\documentclass[epj,nopacs,final]{svjour}
\usepackage[scaled=0.9]{helvet}
\usepackage[utf8]{inputenc}
\usepackage{graphicx}
\usepackage[switch]{lineno}
\usepackage{cite}
\usepackage{subfig}
\usepackage{multirow}
\usepackage{amsmath}
\usepackage{textcomp}
\usepackage{footnote}
\usepackage[compat=1.1.0]{tikz-feynman}
\usepackage{stfloats}
\usepackage[flushmargin]{footmisc}
\usepackage[pdfstartview=XYZ,
bookmarks=true,
colorlinks=true,
linkcolor=blue,
urlcolor=blue,
citecolor=blue,
pdftex,
bookmarks=true,
linktocpage=true, 
hyperindex=true
]{hyperref}

\usepackage{orcidlink}

\newcommand*\patchAmsMathEnvironmentForLineno[1]{%
  \expandafter\let\csname old#1\expandafter\endcsname\csname #1\endcsname
  \expandafter\let\csname oldend#1\expandafter\endcsname\csname end#1\endcsname
  \renewenvironment{#1}%
     {\linenomath\csname old#1\endcsname}%
     {\csname oldend#1\endcsname\endlinenomath}}%
\newcommand*\patchBothAmsMathEnvironmentsForLineno[1]{%
  \patchAmsMathEnvironmentForLineno{#1}%
  \patchAmsMathEnvironmentForLineno{#1*}}%
\AtBeginDocument{%
\patchBothAmsMathEnvironmentsForLineno{equation}%
\patchBothAmsMathEnvironmentsForLineno{align}%
\patchBothAmsMathEnvironmentsForLineno{flalign}%
\patchBothAmsMathEnvironmentsForLineno{alignat}%
\patchBothAmsMathEnvironmentsForLineno{gather}%
\patchBothAmsMathEnvironmentsForLineno{multline}%
}

\journalname{Eur. Phys. J. C}
\title{Measurement of the double-$\boldsymbol{\beta}$ decay of $\boldsymbol{^{150}}$Nd 
to the 0$\boldsymbol{^+_1}$ excited state of $\boldsymbol{^{150}}$Sm in NEMO-3}

\renewcommand{\thefootnote}{\fnsymbol{footnote}}

\author{
  X.~Aguerre\inst{1}
  \and R.~Arnold\inst{2}$^{,\dag}$
  \and C.~Augier\inst{3}
  \and A.S.~Barabash\,\orcidlink{0000-0002-5130-0922}\inst{4}
  \and A.~Basharina-Freshville\inst{5}
  \and S.~Blondel\inst{3}
  \and S.~Blot\inst{6}
  \and M.~Bongrand\inst{3}
  \and R.~Breier\inst{7}
  \and V.~Brudanin\inst{4}$^{,\dag}$
  \and J.~Busto\inst{8}
  \and A.~Bystryakov\,\orcidlink{0000-0002-8212-5098}\inst{4}
  \and A.J.~Caffrey\inst{9}
  \and C.~Cerna\inst{1}
  \and J.P.~Cesar\inst{10}
  \and M.~Ceschia\inst{5}
  \and E.~Chauveau\inst{1}
  \and A.~Chopra\inst{5}
  \and L.~Dawson\inst{5} 
  \and D.~Duchesneau\inst{12}
  \and D.~Durand\inst{11}
  \and J.J.~Evans\inst{6}
  \and R.~Flack\inst{5}
  \and P.~Franchini\inst{13}
  \and X.~Garrido\inst{3}
  \and C.~Girard-Carillo\inst{3}
  \and B.~Guillon\inst{11}
  \and P.~Guzowski\inst{6}
  \and M. Hoballah\inst{3}
  \and R.~Hod\'{a}k\inst{14}
  \and P.~Hubert\inst{1}
  \and M.H.~Hussain\inst{5}
  \and S.~Jullian\inst{3}
  \and A.~Klimenko\,\orcidlink{0000-0003-1993-1094}\inst{4}
  \and O.~Kochetov\,\orcidlink{0009-0001-2327-8334}\inst{4}
  \and S.I.~Konovalov\,\orcidlink{0009-0001-9665-0344}\inst{4}
  \and F.~Ko\v{n}a\v{r}\'ik\inst{14,15}
  \and T.~K\v{r}i\v{z}\'{a}k\inst{14,15}
  \and D.~Lalanne\inst{3}$^{,\dag}$
  \and K.~Lang\,\orcidlink{0000-0003-1269-7223}\inst{10}
  \and Y.~Lemi\`ere\inst{11}
  \and P.~Li \inst{16}
  \and P.~Loaiza\inst{3}
  \and G.~Lutter\inst{1}
  \and M.~Macko\inst{14}  
  \and F.~Mamedov\inst{14}
  \and C.~Marquet\inst{1}
  \and F.~Mauger\inst{11}
  \and A.~Minotti\inst{12}
  \and B.~Morgan\inst{17}
  \and I.~Nemchenok\inst{4}
  \and M.~Nomachi\inst{18}
  \and F.~Nowacki\inst{2}
  \and H.~Ohsumi\inst{19}
  \and G.~Olivi\'ero\inst{11}
  \and V.~Palu\v{s}ov\'{a}\inst{14}
  \and C.~Patrick\,\orcidlink{0000-0002-0713-7515}\inst{16} 
  \and F.~Perrot\inst{1}
  \and M.~Petro\inst{7,14}
  \and A.~Pin\inst{1}
  \and F.~Piquemal\inst{1}
  \and P.~Povinec\inst{7}
  \and S.~Pratt\inst{16}
  \and P.~P\v{r}idal\inst{14}
  \and W.S.~Quinn\inst{5}
  \and Y.A.~Ramachers\inst{17}
  \and A.~Remoto\inst{12}
  \and J.L.~Reyss\inst{20}
  \and C.L.~Riddle\inst{9}
  \and E.~Rukhadze\inst{14}
  \and R.~Saakyan\inst{5}
  \and A.~Salamatin\,\orcidlink{0000-0002-7834-8512}\inst{4}
  \and R.~Salazar\inst{10}
  \and X.~Sarazin\inst{3}
  \and J.~Sedgbeer\inst{13}
  \and Yu.~Shitov\inst{14}
  \and L.~Simard\inst{3,21}$^{,*}$
  \and F.~\v{S}imkovic\inst{7,14}
  \and A.~Smetana\inst{14}
  \and A.~Smolnikov\,\orcidlink{0000-0001-9108-5254}\inst{4}
  \and S.~S\"oldner-Rembold,\orcidlink{0000-0002-9079-6860}\inst{6}\
  \and I.~\v{S}tekl\inst{14}
  \and J.~Suhonen\,\orcidlink{0000-0002-9898-660X}\inst{22}
  \and G.~Szklarz\inst{3}
  \and H.~Tedjditi\inst{8}
  \and J.~Thomas\inst{5}
  \and V.~Timkin\,\orcidlink{0009-0002-4259-5017}\inst{4}
  \and V.I.~Tretyak\,\orcidlink{0000-0002-2369-0679}\inst{23,24}
  \and V.I.~Tretyak\,\orcidlink{0000-0002-0294-4174}\inst{4}
  \and V.I.~Umatov\,\orcidlink{0000-0001-6881-3540}\inst{4}
  \and I.~Vanushin\inst{4}$^{,\dag}$
  \and Y.~Vereshchaka\inst{3}
  \and V.~Vorobel\inst{25}
  \and D.~Waters\inst{5}
  \and F.~Xie\inst{5} 
}
\institute{
  Universit\'e de Bordeaux, CNRS/IN2P3, LP2i Bordeaux, UMR 5797, F-33170 Gradignan, France
  \and Universit\'e Louis Pasteur, CNRS/IN2P3, IPHC, F-67037 Strasbourg, France
  \and Universit\'e Paris-Saclay, CNRS, IJCLab, F-91405 Orsay, France
  \and Participant in the NEMO-3/SuperNEMO collaboration
  \and University College London, London, WC1E 6BT, United Kingdom
  \and University of Manchester, Manchester, M13 9PL,~United Kingdom
  \and Faculty of Mathematics, Physics and Informatics, Comenius University, SK-842 48 Bratislava, Slovakia
  \and Aix-Marseille Universit\'e, CNRS, CPPM, F-13288 Marseille, France
  \and Idaho National Laboratory, Idaho Falls, ID 83415, U.S.A.
  \and University of Texas at Austin, Austin, TX 78712, U.S.A.
  \and Normandie Universit\'e, ENSICAEN, UNICAEN, CNRS/IN2P3, LPC Caen, F-14000 Caen, France
  \and Universit\'e de Savoie, CNRS/IN2P3, LAPP, UMR 5814, F-74941 Annecy-le-Vieux, France
  \and Imperial College London, London, SW7 2AZ, United Kingdom
  \and Institute of Experimental and Applied Physics, Czech Technical University in Prague, CZ-11000 Prague, Czech Republic
  \and Faculty of Nuclear Sciences and Physical Engineering, Czech Technical University in Prague, Brehova 7, 115 19 Prague, Czech Republic
  \and University of Edinburgh, Edinburgh, EH9 3FD, United Kingdom  
  \and University of Warwick, Coventry, CV4 7AL, United Kingdom
  \and Osaka University, 1-1 Machikaneyama Toyonaka, Osaka 560-0043, Japan
  \and Saga University, Saga 840-8502, Japan
  \and LSCE, CNRS, F-91190 Gif-sur-Yvette, France
  \and Institut Universitaire de France, F-75005 Paris, France
  \and Jyv\"askyl\"a University, FIN-40351 Jyv\"askyl\"a, Finland
  \and Institute for Nuclear Research of NASU, 03028 Kyiv, Ukraine
  \and INFN - Laboratori Nazionali del Gran Sasso, 67100 Assergi (AQ), Italy
  \and Charles University in Prague, Faculty of Mathematics and Physics, CZ-12116 Prague, Czech Republic
}

\date{Received: date / Accepted: date}
\abstract{
The NEMO-3 results for the double-$\beta$ decay of $^{150}$Nd
to the 0$^+_1$ and 2$^+_1$ excited states of $^{150}$Sm are reported. 
The data recorded during 5.25 yr with 36.6 g of the isotope $^{150}$Nd
are used in the analysis.
For the first time, the signal of the 
$2\nu\beta\beta$ transition to the 0$^+_1$ excited state is detected with a statistical significance exceeding 5$\sigma$.
The half-life is measured to be
$T_{1/2}^{2\nu\beta\beta}(0^+_1) = \left[ 1.11  ^{+0.19}_{-0.14} \,\left(\mbox{stat}\right) ^{+0.17}_{-0.15}\,\left(\mbox{syst}\right) \right] \times10^{20}$~yr. 
The limits are set on the $2\nu\beta\beta$ decay to the 2$^+_1$ level 
and on the $0\nu\beta\beta$ decay to the 0$^+_1$ and 2$^+_1$ levels 
of $^{150}$Sm.
}
\begin{document}
\maketitle
\renewcommand{\thefootnote}{\fnsymbol{footnote}}
\footnotetext[1]{{\it{Correspondence to}}: laurent.simard@ijclab.in2p3.fr}
\footnotetext[2]{Deceased}
\section{Introduction}
\label{Introduction}
The double-$\beta$ decay is a nuclear process that changes the charge of a nucleus by two units through the simultaneous $\beta$-decay of two constituent neutrons
to protons. 
The two-neutrino double-$\beta$ decay ($2\nu\beta\beta$)  is a rare second-order weak interaction process
occurring with emission of two electrons and two antineutrinos. It was observed for several nuclear isotopes ~\cite{Zyla:2020zbs},~\cite{universe6100159}.

The community's interest in the double-$\beta$ decay is particularly motivated by the search for its 
hypothetical neutrinoless mode ($0\nu\beta\beta$)~\cite{0nu-review}. This process violates the lepton number conservation and is only possible if the neutrino has mass and is a 
Majorana particle~\cite{Majorana}, i.e. $\nu \equiv \bar\nu$. The discovery of $0\nu\beta\beta$ would indicate the physics beyond the Standard Model (BSM).

The rates of two-neutrino and neutrinoless double-$\beta$ decay may be expressed
as 
\begin{equation}
1/T^{2\nu}_{1/2} = G^{2\nu} {g}^4_A |M^{2\nu}|^2~, 
\label{eq:one}
\end{equation}
\begin{equation}
1/T^{0\nu}_{1/2} = G^{0\nu} {g}^4_A  \vert M^{0\nu} \vert ^{2} \langle \eta \rangle^{2} \,,
\label{eq:two}
\end{equation}
where $G^{2\nu,0\nu}$ are the phase space factors, $g_A$ is the axial
vector coupling constant, $M^{2\nu,0\nu}$ are the nuclear matrix elements (NMEs) for the corresponding decay modes, and $\langle \eta \rangle$ is 
a parameter of the underlying BSM physics model 
(in the case of the commonly considered mass mechanism of the $0\nu\beta\beta$ decay, the exchange of a light Majorana neutrino,
 $\langle \eta \rangle$ is 
the effective neutrino mass).
The phase space factors can be accurately calculated while the model-dependent NME calculations have a substantial theoretical uncertainty.
The measurement of the $2\nu\beta\beta$ decay half-life provides valuable information 
for nuclear structure models used in NME calculations.

The double-$\beta$ decay can proceed through transitions either to the ground state or to excited states of the daughter nucleus. The latter occurs at a lower rate because of its smaller transition energy $Q_{\beta\beta}$ leading to the correspondingly suppressed phase space factor.
Nevertheless, the measurement of the $2\nu\beta\beta$ decay to excited states
provides supplementary information for nuclear models.
Additionally, in the case of the $0\nu\beta\beta$ decay discovery, the ratio of half-lives for
transitions to the 0$^+$ first excited state and the ground state may allow the dominant decay mechanism to be determined ~\cite{Simkovic:2001ft}.
Information on the results of experiments on the $\beta\beta$-decay to excited states of
daughter nuclei can be found in~\cite{universe6120239}.

The isotope $^{150}$Nd is one of the best candidates for  neutrinoless $\beta\beta$-decay searches
because of its high transition energy $Q_{\beta\beta}$ = 3371\,keV and highest phase space factor~\cite{10.3389/fphy.2019.00012}. However, a modest isotopic abundance of 5.638(28)\%  and difficulties in isotopic enrichment  limit its use in large-scale experiments.

\begin{figure}
\begin{tikzpicture}[xscale=0.089,yscale=0.1,>=stealth ]
\draw (29,70) -- node[below] {$^{150}_{~61}\mbox{Pm}$} (49,70);
\draw (5,60) -- node[below] {$^{150}_{~60}\mbox{Nd}$} (25,60)
node at(25,62.5) {$0^+$};
\draw (54,32) --  (79,32)
node at(87,34.5) {$0^+_1$   740 keV};
\draw  [->, thin] (57,32) --  node[right] {406 keV} (57,20);
\draw (54,20) --  (79,20);
\draw  [->, thin] (60,20) --  node[right] {334 keV} (60,10)
node at(87,22.5) {$2^+_1$   334 keV};
\draw (54,10) -- node[below] {$^{150}_{~62}\mbox{Sm}$} (79,10)
node at(81,12.5) {$0^+_{\textsf{g.s.}}$};
\draw  [->, thin] (25,60) -- (54,32) 
 node at (45,45) {$\beta\beta$};
\draw  [->, thin] (25,60) -- (54,20);
\draw  [->, thin] (25,60) -- (54,10)
node at (20,10) {Q$_{\beta\beta}$=3371~keV};
\end{tikzpicture}
\caption{Scheme of the $^{150}$Nd $\beta\beta$-decay to the lowest excited states
of $^{150}$Sm}
\label{fig:scheme}
\end{figure}
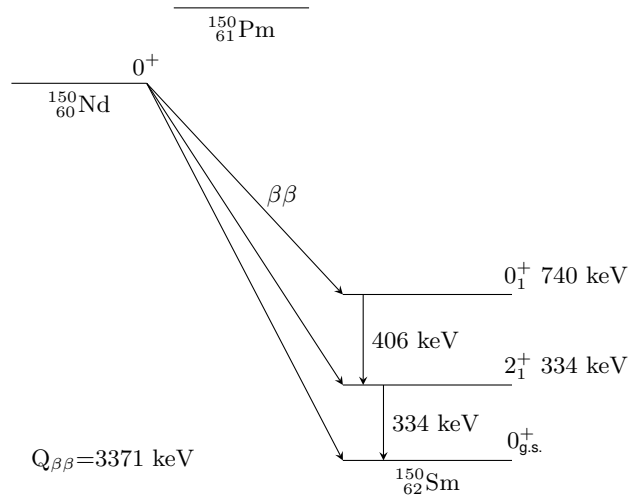

The decay scheme of $^{150}$Nd to the 2$^+_1$ and 0$^+_1$ excited states is shown in Fig.~\ref{fig:scheme}.
The half-life for the transition to the first 0$^+$
excited state was first measured in 2004~\cite{BarHubHub04}.
These data were subsequently re-analysed, with the final result published in \cite{PhysRevC.79.045501}.
The measurements of this decay were obtained with $\gamma$-ray spectrometry  using high-purity
germanium detectors \cite{PhysRevC.79.045501},\cite{PhysRevC.90.055501},\cite{Polischuk}; 
none of the previous measurements detected a signal with a 
5$\sigma$ statistical significance.
 For other excited states, only lower limits on the half-life were established; the 
best available limit ~\cite{PhysRevC.79.045501} for the transition to the $2^+_1$ state is 
$T_{1/2}^{\beta\beta}(2^+_1) > 2.2\times10^{20}\,\mbox{yr}$.

The most precise measurement for the $\beta\beta$-decay of $^{150}$Nd  to the ground state 
was performed by NEMO-3~\cite{Arnold:2016nd150}:
\begin{equation}
T_{1/2}^{2\nu\beta\beta}(0^+_{\textsf{g.s.}}) = [9.34\pm 0.22 \,(\mbox{stat}) ^{+0.62}_{-0.60}(\mbox{syst})] \times10^{18}\,\mbox{yr}.
\label{eq:three}
\end {equation}
It is based on the 
data recorded for 5.25 years with 36.6~g of $^{150}$Nd. The same data set is used
in this analysis. 
\section{NEMO-3 detector}
\label{detector}
The NEMO-3 experiment  in the Modane Underground Laboratory (LSM)  took data
from February 2003 to January 2011. The NEMO-3 detector, designed 
to search for the $0\nu\beta\beta$ decay,
uses both a tracking device and a calorimeter, which enables the
direct 
detection of electrons, positrons, photons, and $\alpha$-particles.

\begin{figure}
\includegraphics[width=0.49\textwidth]{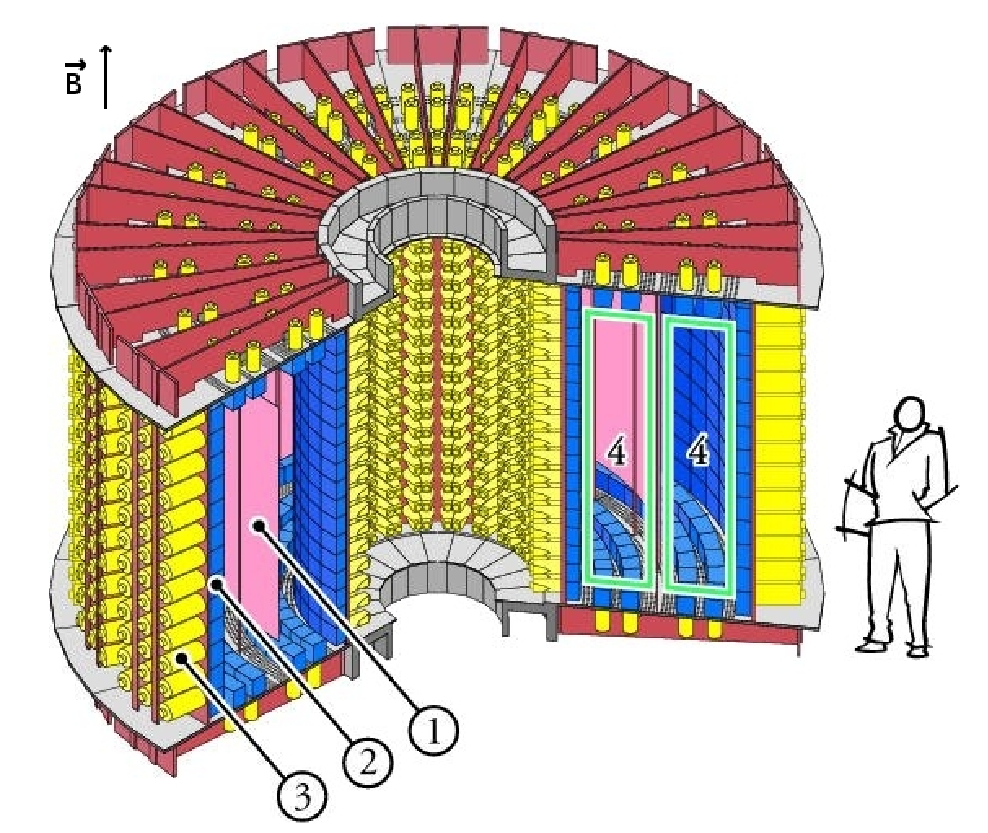}
\caption{Schematic view of the NEMO-3 detector with the source foils (1), scintillators (2),
photomultipliers (3), and wire chamber (4)}
\label{fig:scheme_NEMO-3}
\end{figure}

A schematic view of the NEMO-3 detector is shown in Fig.~\ref{fig:scheme_NEMO-3}.
The detector was a hollow cylinder with a diameter of 5~m
and a height of 3~m and was composed of 20 equal sectors.
These hosted thin source foils of 7 different enriched $\beta\beta$-decaying isotopes
($^{100}$Mo, $^{82}$Se, $^{116}$Cd, $^{130}$Te, $^{150}$Nd, $^{96}$Zr, and $^{48}$Ca) 
with a total mass of about 10~kg.

The source foils were suspended vertically 
between two concentric cylindrical tracker volumes, parallel to the wires of the tracking detector. The tracking detector  
was composed of 6180 open octagonal drift cells arranged in 18 concentric layers,
with 9 layers in each of the two volumes. The tracker was filled with 
a gas mixture of helium (94.9\%), ethyl alcohol (4\%),  argon (1\%), and  water vapour (0.1\%)
at 7 mbar above atmospheric pressure. The drift cells operating in the Geiger mode
enabled three-dimensional
measurements of trajectories and decay vertices of charged particles.
The average Geiger cell resolution 
was  0.5 mm in the horizontal plane  and  8 mm
in the vertical direction (parallel to the wires).

The tracking chamber was surrounded by a calorimeter composed of 1940
plastic scintillator blocks coupled to low-radioactivity 3-inch and 5-inch photomultiplier tubes (PMTs). The calorimeter provided both time and energy measurements. The energy
resolution of the calorimeter was $\sigma$ = (5.8 -- 7.2)\%, 
and the time resolution 
was $\sigma$~=~250~ps for 1-MeV electrons.

A vertical magnetic field of 25 Gauss inside the wire chamber was provided by a solenoidal coil.
The detector was surrounded by the passive shielding consisting of  19-cm-thick iron plates  to 
suppress the external
$\gamma$-ray flux and also of borated water, paraffin, and wood to thermalize
and absorb environmental neutrons.
The experimental hall is located at a depth of 4800\,m.w.e., to reduce the cosmic-ray flux.

The $^{150}$Nd foil was manufactured using Nd$_2$O$_3$ powder provided by the Institute for Nuclear Research of RAS in Moscow. 
Neodymium was enriched by 
electromagnetic separation to (91.0$\pm$0.5)\% of the isotope $^{150}$Nd  and chemically purified. A total of 46.64 g of Nd$_2$O$_3$ powder
mixed with a concentration of 8\% PVA glue  was uniformly distributed between two layers of
mylar to produce a composite foil with a total mass of 
56.68~g.
The foil
was 2484~mm long and 65 mm wide. 
The total mass of 
the isotope $^{150}$Nd in the foil was 36.6$\pm$0.2~g ~\cite{Arnold:2004TDR}.
The $^{150}$Nd composite foil was located in Sector 5 of the NEMO-3 detector between a foil of $^{100}$Mo and a
foil containing $^{96}$Zr and $^{48}$Ca.

A more detailed description of the NEMO-3 detector, its calibration and performance can be found in~\cite{Arnold:2004TDR} and ~\cite{Arnold:2015wpy}.
\section{Analysis and results}
The $\beta\beta$-decay of $^{150}$Nd to the lowest (2$^+_1$ and 0$^+_1$)
excited states of $^{150}$Sm
has been investigated. The contribution from the higher excited states was 
neglected.
According to the decay scheme in Fig.~\ref{fig:scheme}, 
two electrons from the $\beta\beta$-decay 
are accompanied by one $\gamma$ in the case of the transition
to the 2$^+_1$ excited state  and by 
two photons in the transition to the 0$^+_1$ excited state. 
We therefore select for this analysis two-electron one-$\gamma$ ($ee\gamma$) and two-electron two-$\gamma$ ($ee\gamma\gamma$) event topologies.
After the event selection,  the $\beta\beta$-decay signal
is identified by an excess in the data over the 
expected background. 

Both a measurement of the two-neutrino $\beta\beta$-decay and a search for the neutrinoless $\beta\beta$-decay
to the 0$^+_1$ excited state are carried out in the $ee\gamma\gamma$
and $ee\gamma$ channels. The $\beta\beta$-decays to the
2$^+_1$ excited state  are explored in the $ee\gamma$ channel. 

A multivariate analysis improves the separation 
of the signal from the background.
To this end the Boosted Decision Tree (BDT) method 
is used. 
The analysis employs a BDT algorithm with adaptive boosting, part of the ROOT~\cite{ROOT}  TMVA 
package~\cite{TMVA}.

Where no evidence of a signal in the data is found, a limit
on the corresponding decay half-life is set. The 90\% confidence level (C.L.)
limit is calculated using the CL$_s$ method 
employing the modified frequentist approach ~\cite{Junk:1999kv,Read,Fisher:2006zz}.

\subsection{Event selection}
In this analysis, the $ee\gamma\gamma$ and $ee\gamma$ event topologies
are used.
Events are selected by requiring two reconstructed electron tracks coming from the source foil, with
each depositing in a separate scintillator block an energy greater than 150\,keV.
Extrapolating each track to the source foil gives the position of its decay vertex,
and extrapolating to the calorimeter associates the track with the scintillator block of an optical module  for energy and time measurement.
A scintillator hit associated with a track must be isolated, i.e.
no hits should be found 
in neighboring scintillator blocks. 

Each of the two electron tracks must have a length greater than 50\,cm 
and originate from a
common vertex in the $^{150}$Nd source foil: the separation between
the two individually reconstructed track vertices is required to be less than 4\,cm 
in the horizontal plane and less than 8\,cm in the vertical direction. 
An event is excluded if its vertex is found in one of the regions of
the enhanced activity in the foil corresponding to the localized contamination
from $^{234m}$Pa and $^{207}$Bi (hot spots).
The locations of the hot spots, which amount to 7\% of the $^{150}$Nd foil area, were determined in~\cite{Arnold:2016nd150}.

To ensure that an event corresponded to the simultaneous emission of two electrons from a common vertex, the 
corresponding time-of-flight (TOF) probability is required to be higher than 5\%.
The TOF probability is calculated using energy and time 
measurements from the calorimeter and the 
distances travelled by particles in the event; see \cite{Arnold:2004TDR} and \cite{Arnold:2015wpy} 
for details.

A $\gamma$-ray is identified as either a single calorimeter hit or a cluster 
of neighbouring hits that are not associated with any track. 
A minimum threshold of 100\,keV for the
energy deposited in each of these calorimeter blocks is required.
It is also required that no prompt Geiger hits are detected within 20\,cm of 
any scintillator block attributed to a $\gamma$-ray.
Events are rejected if the TOF probability exceeds 1\% for the hypothesis
that the event originates from an external $\gamma$-ray. 
The probability for the hypothesis that the photon(s) originated
from the event vertex simultaneously with two electrons is required 
to be higher than 5\%.

An event is rejected if it contains a recognized delayed alpha-particle track,
as described in~\cite{Arnold:2015wpy},
to reduce the background from
$^{214}$Bi decays.

As shown in Table~\ref{table:bkg_all},
a total of 142 $ee\gamma\gamma$ and 571 $ee\gamma$ events are selected 
from the full data set.
\begingroup
\renewcommand{\arraystretch}{1.2} 
\begin{table}[hbt]
\begin{center}
\caption{Expected number of events from different sources of the background
with statistical and systematic uncertainties 
and the number of the observed events in the $ee\gamma\gamma$ and $ee\gamma$
channels after the event selection}
\label{table:bkg_all}
\begin{tabular}{ l |  l | l }
\hline \hline
Contribution & $ee\gamma\gamma$ & $ee\gamma$  \\
\hline
$^{228}$Ac+$^{212}$Bi+$^{208}$Tl  & 65.81$\pm$0.39$\pm$4.61          & 279.0$\pm$0.9$\pm$19.5 \\
$^{214}$Bi                        & 7.33$\pm$0.08$\pm$1.69           &  48.5$\pm$0.2$\pm$11.2    \\
$^{152}$Eu+$^{154}$Eu             & 3.57$\pm$0.08$\pm$0.42           &  36.5$\pm$0.3$\pm$5.1      \\
$^{207}$Bi                        & 1.78$\pm$0.06$\pm$0.10           &  41.2$\pm$0.3$\pm$2.3     \\
$^{234m}$Pa                       & 0.02$\pm$0.02$\pm$0.002          &   4.9$\pm$0.3$\pm$0.5      \\
Radon                             & 3.26$\pm$0.11$\pm$0.33           &  23.3$\pm$0.3$\pm$2.3      \\
External background                      & 2.74$\pm$0.54$^{+0.85}_{-0.63}$  &  47.0$\pm$2.4$^{+14.6}_{-10.8}$    \\
Neighbouring foils                   & 0.59$\pm$0.03$\pm$0.14           &   4.6$\pm$0.5$\pm$1.0      \\
$^{150}$Nd $\beta\beta \to$ g.s.  & 0.25$\pm$0.02$\pm$0.02           &  27.4$\pm$0.2$\pm$1.9      \\
\hline
Total  bkg                        & 85.35$\pm$0.69$^{+5.01}_{-4.98}$& 512.5$\pm$2.7$^{+27.6}_{-25.8}$ \\
\hline
Data                              &   142                           &  571 \\
\hline
\hline
\end{tabular}
\end{center}
\end{table} 
\endgroup
\begin{figure*}[h!]
\begin{center}
\includegraphics[width=0.32\textwidth]{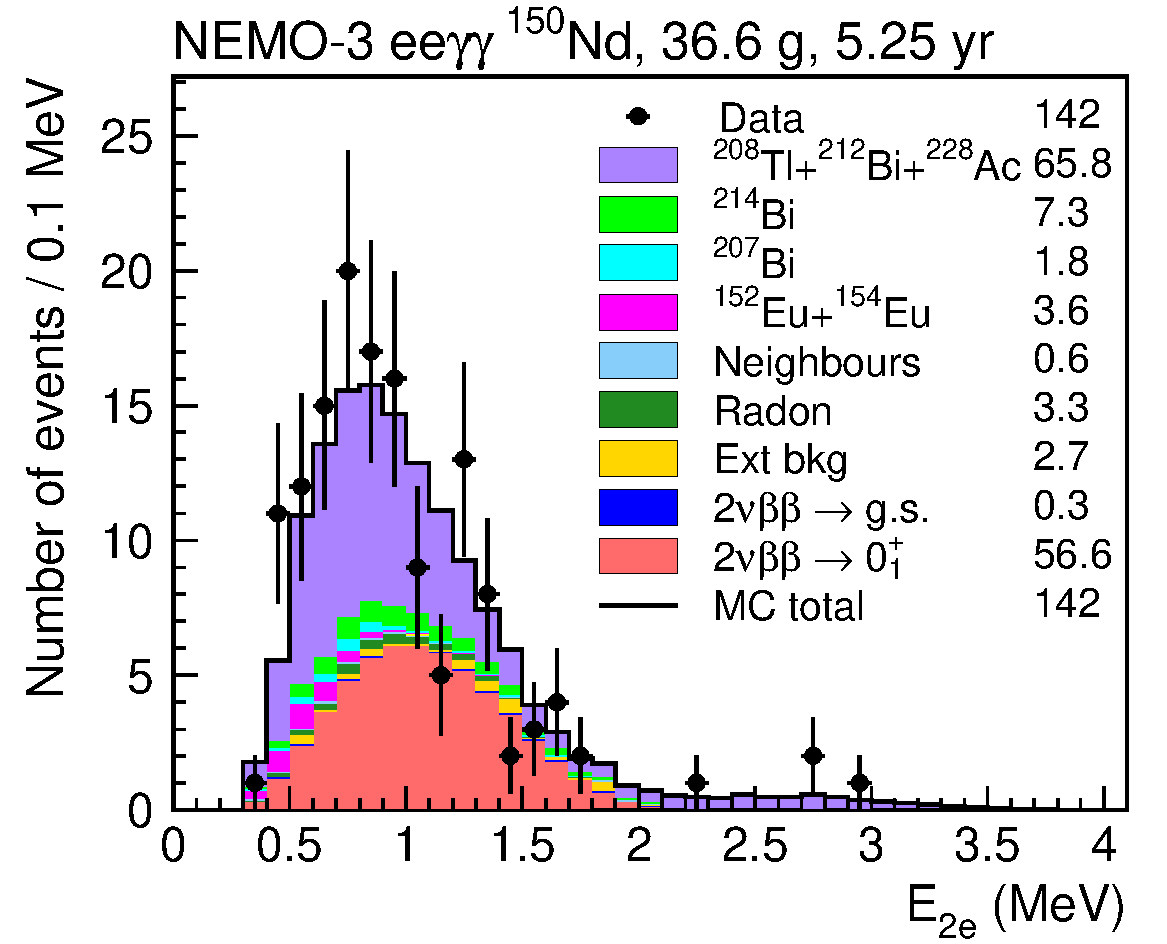}
\includegraphics[width=0.32\textwidth]{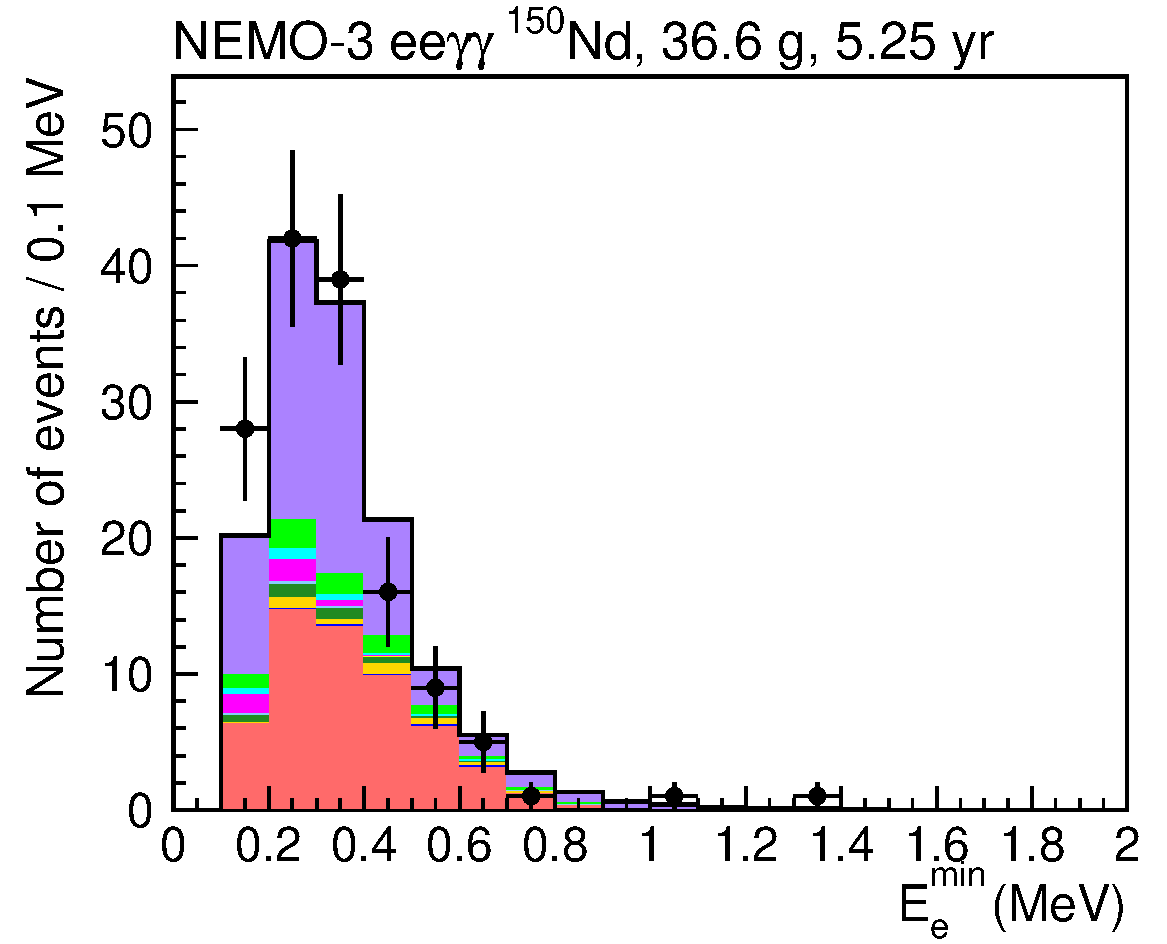}
\includegraphics[width=0.32\textwidth]{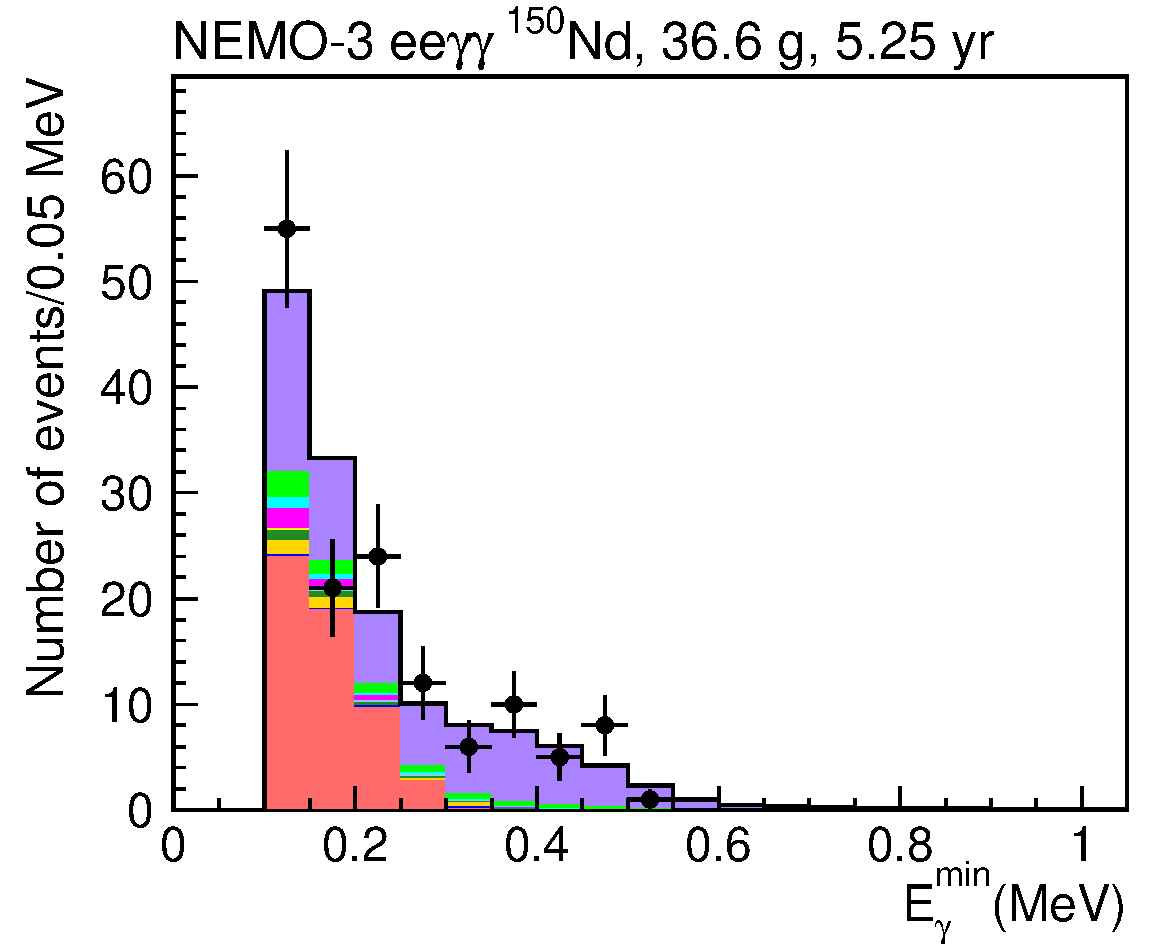}
\includegraphics[width=0.32\textwidth]{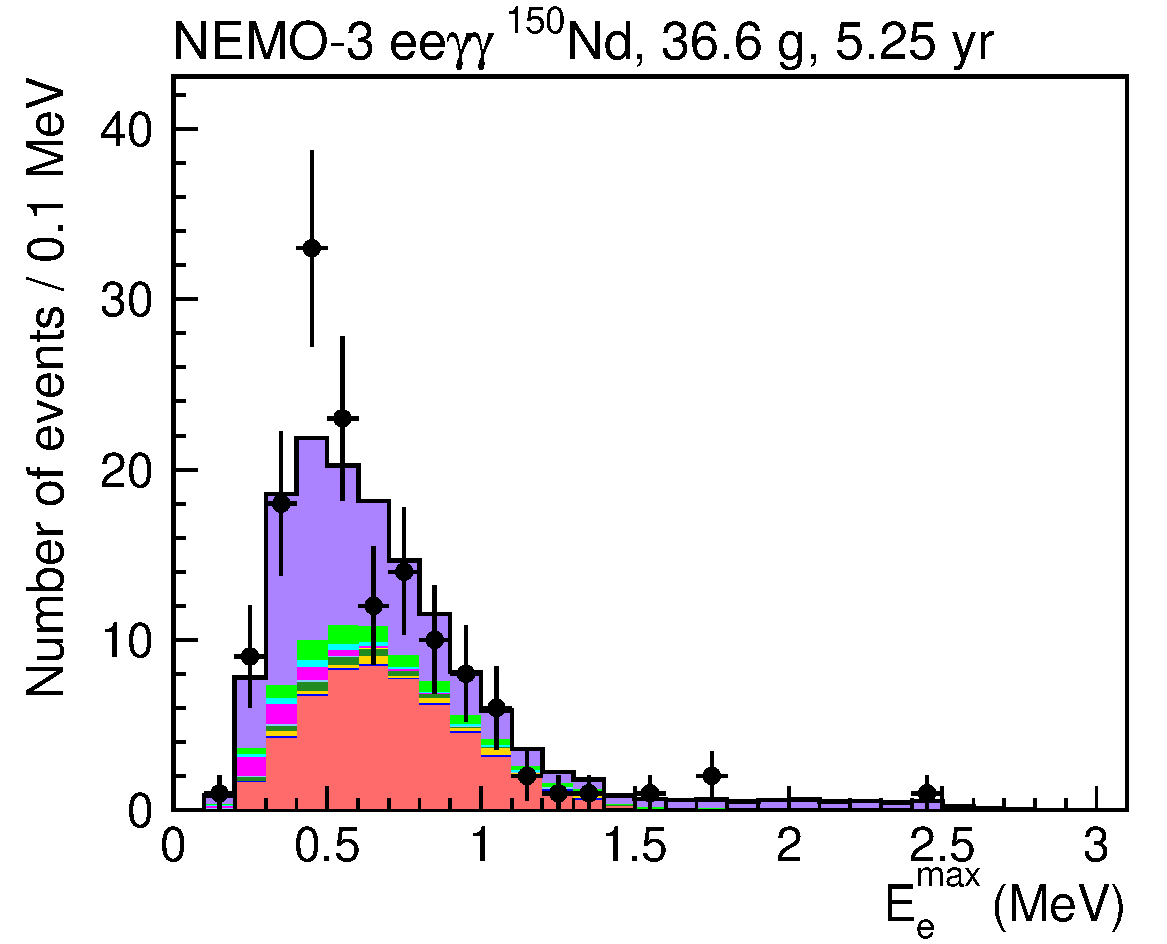}
\includegraphics[width=0.32\textwidth]{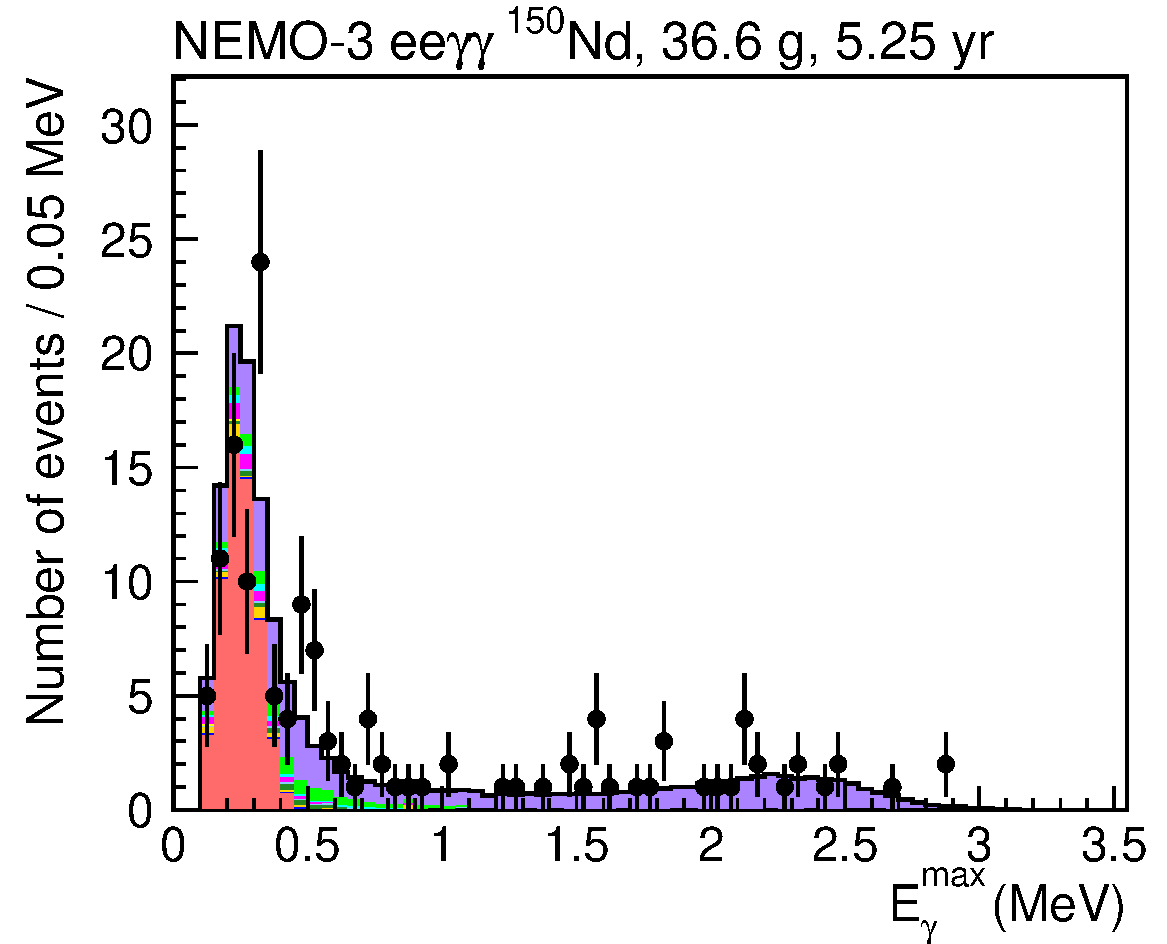}
\includegraphics[width=0.32\textwidth]{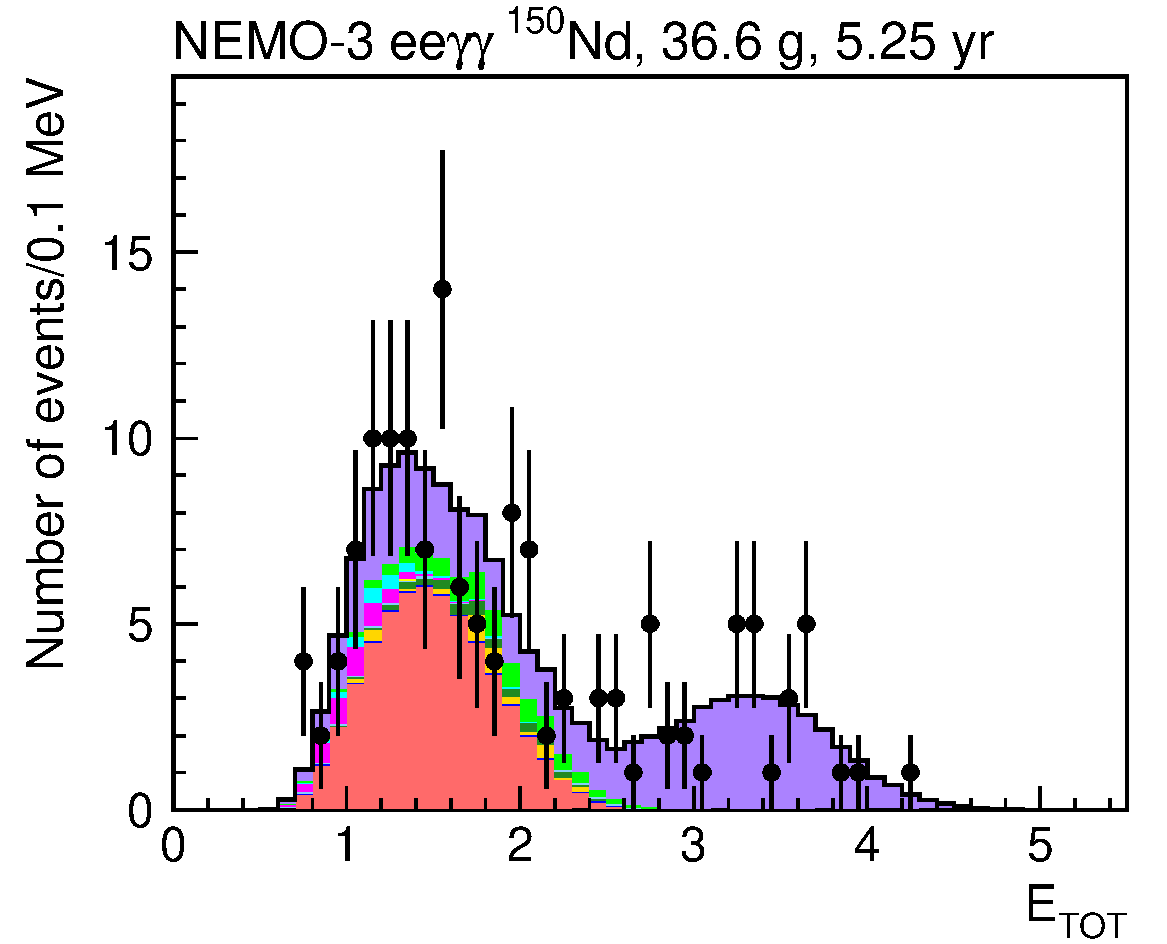}
\includegraphics[width=0.32\textwidth]{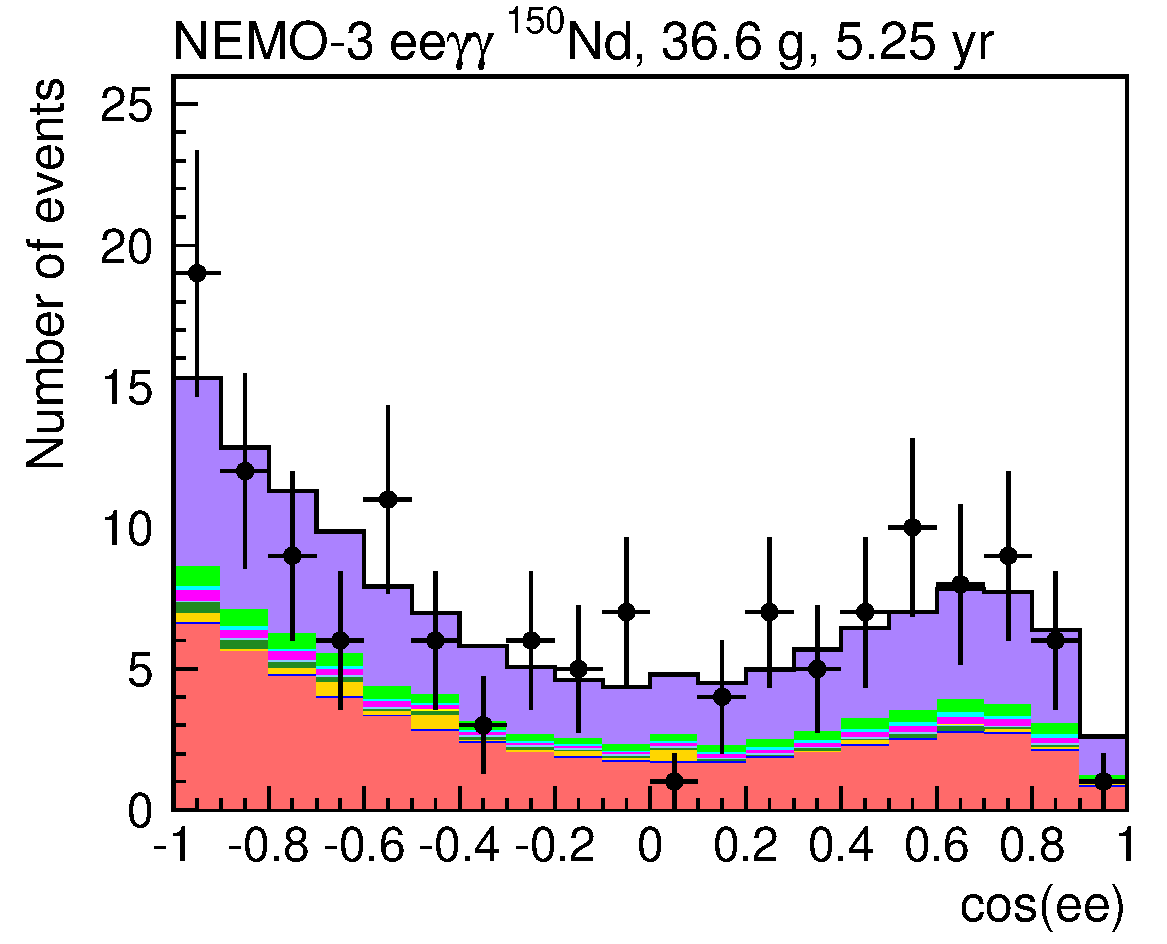}
\includegraphics[width=0.32\textwidth]{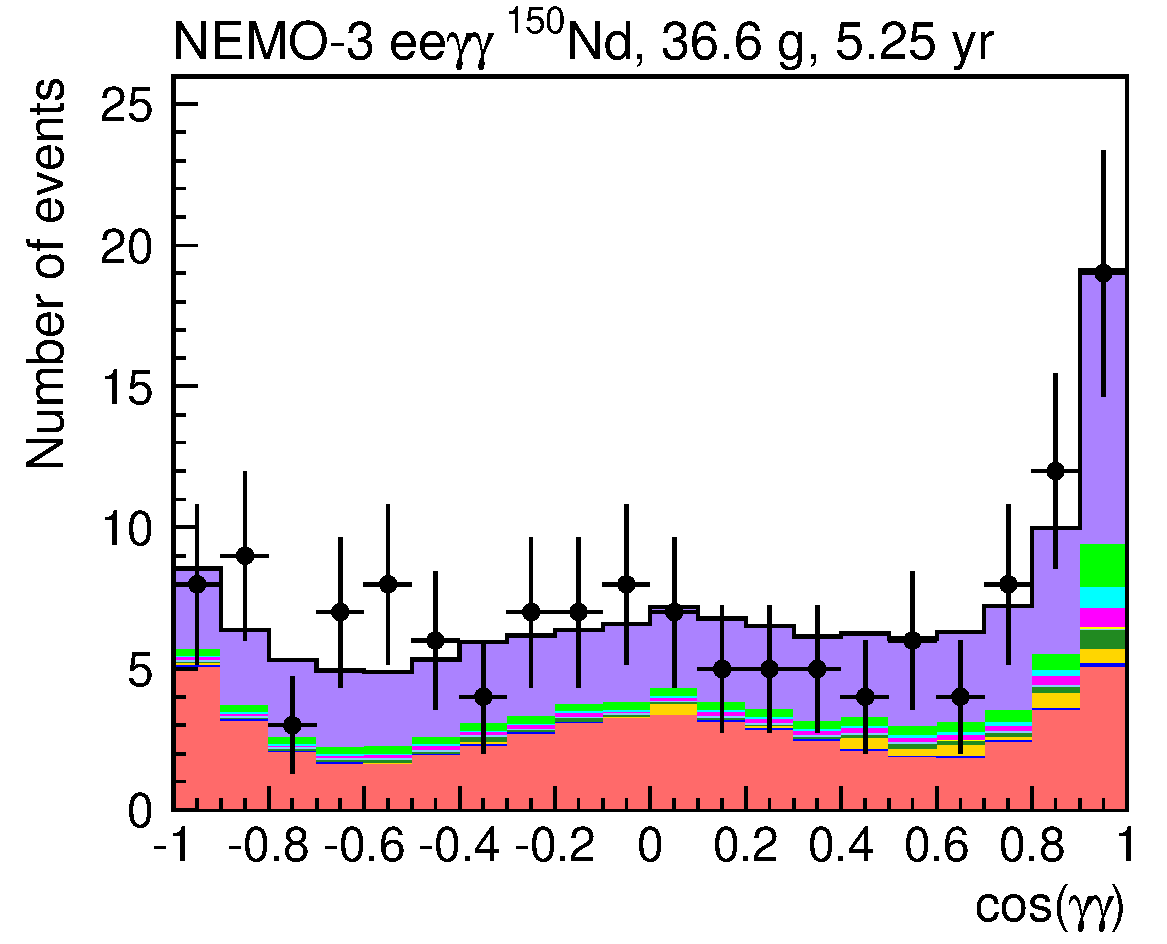}
\includegraphics[width=0.32\textwidth]{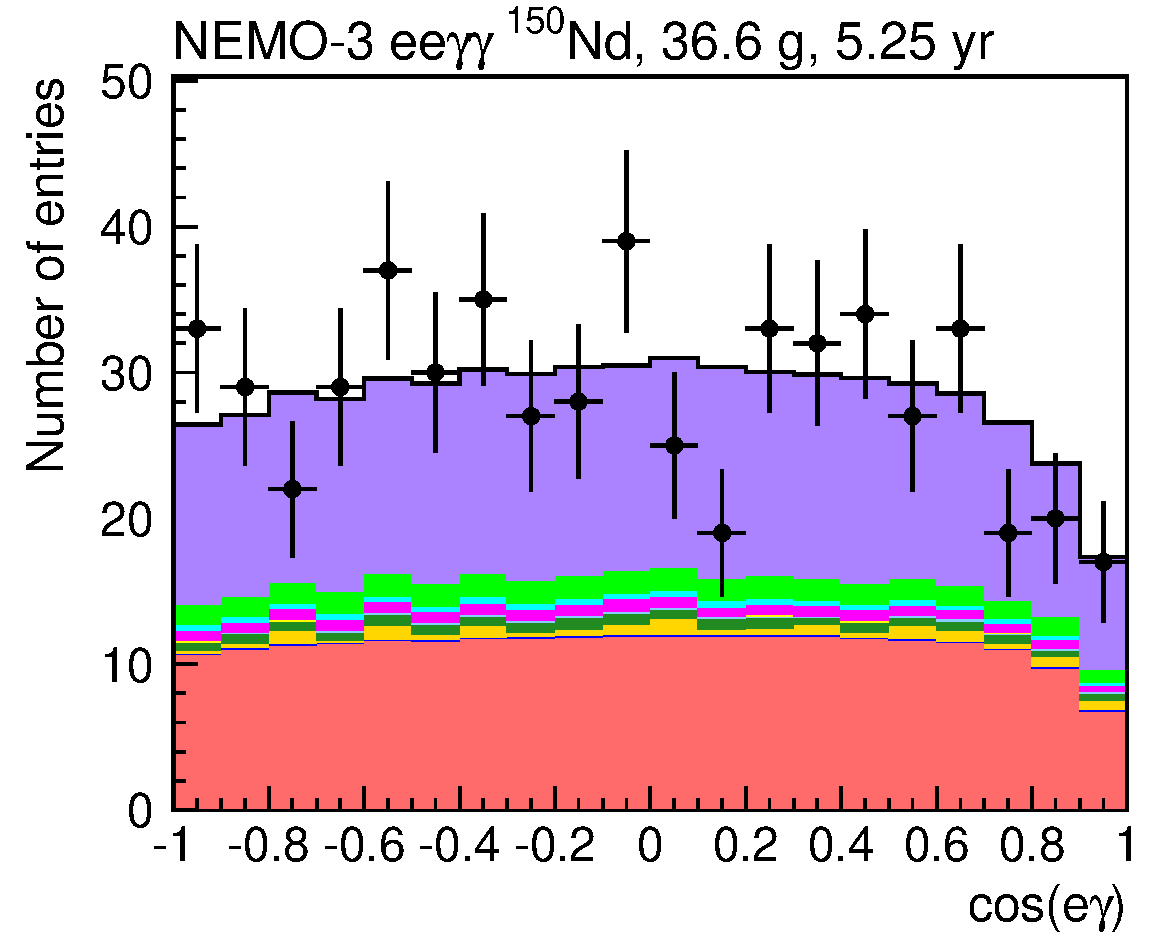}
\caption{Distributions of the measured quantities for the $ee\gamma\gamma$ events from the $^{150}$Nd foil
after the preliminary event selection: energy sum of two electrons $E_{2e}$, minimal electron energy $E_{e}^{\text{min}}$, minimal $\gamma$ energy $E_{\gamma}^{\text{min}}$, 
maximal electron energy $E_{e}^{\text{max}}$, maximal $\gamma$ energy $E_{\gamma}^{\text{max}}$, 
total measured energy $E_\text{TOT}$, cosine of the angle between two electrons $\cos(ee)$, between two photons 
$\cos(\gamma\gamma)$, and between electron and $\gamma$ $\cos(e\gamma)$ for all 
$e\gamma$ combinations.
Data are compared to the MC prediction 
with the resulting number of $0^{+}_{1}$ signal events obtained by background subtraction.}
\label{fig:eegg-prelim2}
\end{center}
\end{figure*}
\subsection{Background model}
The main source of background events is
trace amounts of naturally occurring radioactive isotopes
that come from the $^{238}$U and $^{232}$Th radioactive series.
The most important of them are  ($\beta,\gamma$)-emitting 
isotopes with high $Q_\beta$ values, such as $^{208}$Tl ($Q_\beta$ = 4.99~MeV) 
and $^{214}$Bi ($Q_\beta$ = 3.27~MeV).
According to their origin with respect to the source foil, the background events are 
classified as internal or external ones.

The largest background contribution comes from the internal
contamination of the source foil.
The decay of a $\beta$-emitting isotope inside the foil
can mimic the $\beta\beta$-decay signal through several different 
mechanisms, such as a single $\beta$-decay 
combined with M{\o}ller scattering  or a single $\beta$-decay to an 
excited state of the
daughter nucleus followed by the emission of a conversion
electron or a $\gamma$-ray that undergoes Compton scattering in
the foil.
From these mechanisms, additional $\gamma$-rays could be produced
by bremsstrahlung or from a decay to an excited state.
In addition to the radioactive impurities, a decay in a neighbouring NEMO-3 source
foil can be misinterpreted to have 
its vertex in the $^{150}$Nd foil.
The $^{150}$Nd $\beta\beta$-decay to the ground state also contributes to the background for the excited-state measurement; 
 two electrons are produced in the decay, and
one or two $\gamma$-rays could be emitted via bremsstrahlung.

The external background is there due to the radioactivity 
outside of the source foil. 
Radioactive decays 
within the detector components (mainly PMT glass), the shielding and rock, surrounding the laboratory, generate the external $\gamma$-ray flux.
$\gamma$-ray interactions with the source foil can cause electron–positron pair
production, a Compton interaction followed by M{\o}ller scattering, 
or double Compton scattering. In the case of electron–positron pair production, 
 two photons can be produced through annihilation of the positron.

A subset of the external background is induced by radon.
Radon is a highly diffusive gas and is outgassed into the air from the rock walls of the LSM laboratory.
It is present in the tracker volume due to diffusion from laboratory air through detector seals
and emanation from detector materials.
The decay of radon progenies (mainly $^{214}$Bi) near the
source foil can produce signal-like events similar
to internal background decays.

Details of the background model and measured values of activities that are used in this analysis are provided in~\cite{Arnold:2016nd150}.
The DECAY0 event generator~\cite{Ponkratenko:2000um} is used to simulate
the signal and backgrounds, and particles are tracked through a detailed GEANT3-based detector simulation~\cite{Brun:1987GEANT3}.
Both the data and Monte Carlo (MC) events are processed by
the same reconstruction and selection algorithm.
The number of the expected background events with the $ee\gamma\gamma$ and $ee\gamma$
topologies is given in Table~\ref{table:bkg_all}.
\subsection{Measurement of $\boldsymbol{2\nu\beta\beta}$ decay to $\boldsymbol{0^+_1}$ excited state}
An excess in the data over the total expected background is observed both in the 
$ee\gamma\gamma$ and $ee\gamma$ channels (see Table~\ref{table:bkg_all}) 
and can be attributed to the signal of the $\beta\beta$-decay 
to the excited states of the daughter nucleus $^{150}$Sm.
\subsubsection{Use of $\boldsymbol{ee\gamma\gamma}$ events}
The $ee\gamma\gamma$ event topology is the best one for measuring the transition to the
0$^+_1$ excited state when both electrons, produced in the $\beta\beta$-decay, and both photons from 
deexcitation of $^{150}$Sm are detected.
\begin{figure}[h]
\includegraphics[width=0.48\textwidth]{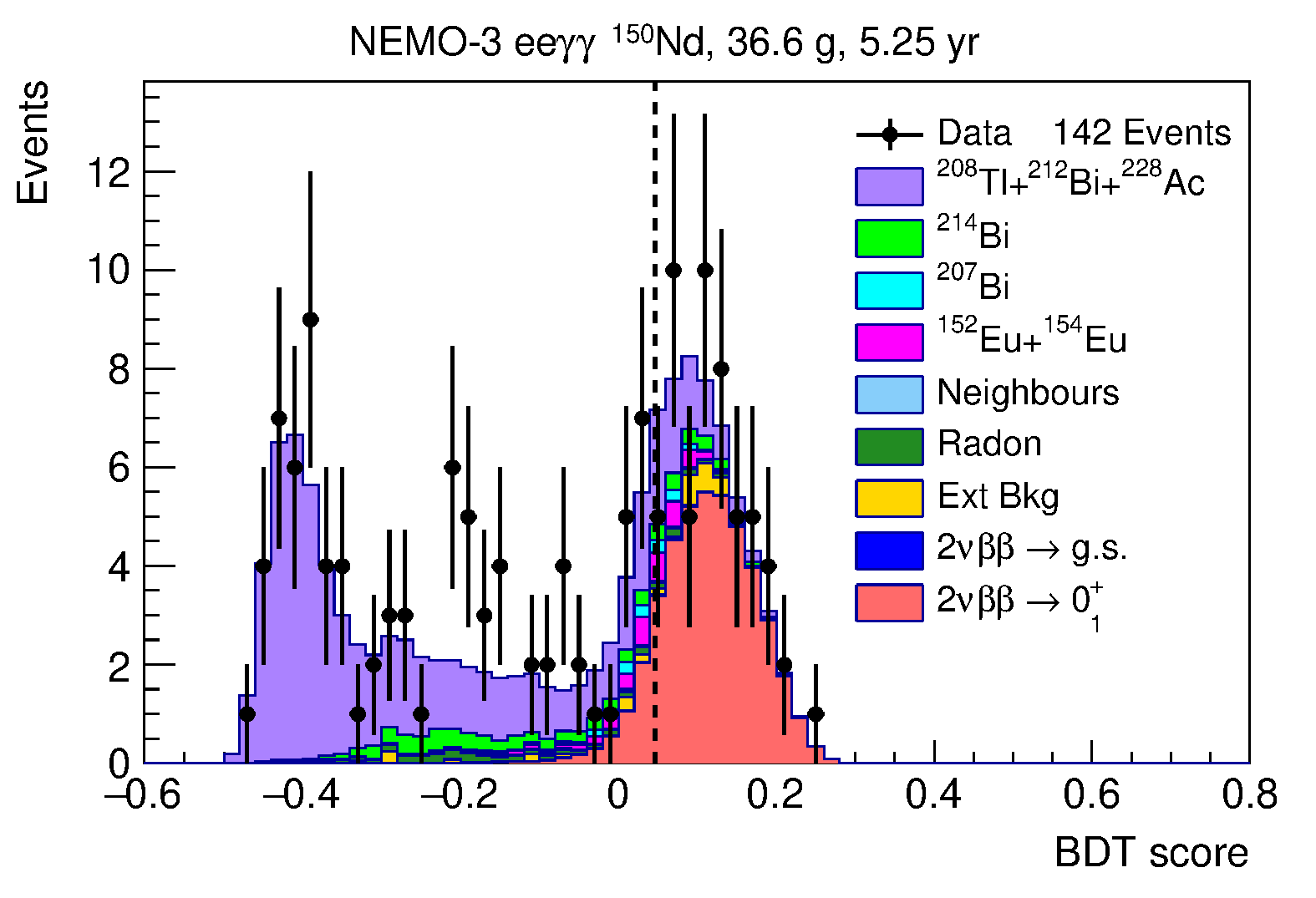}
\caption{BDT score distribution for the $ee\gamma\gamma$ events used for the signal of the $^{150}$Nd $2\nu\beta\beta$ decay to the 0$^+_1$ excited state. The vertical dashed line denotes the optimal cut position maximizing the signal significance.}
\label{fig:bdt_eegg_2nu_0plus1}
\end{figure}
Distributions of measured quantities for the selected $ee\gamma\gamma$ events are demonstrated
in Fig.~\ref{fig:eegg-prelim2}. The number of the 0$^+_1$ signal events $S = N - B = 56.6 \pm 11.9$ is obtained by
subtracting the expected background from the number of events observed. This corresponds to the signal-to-background ratio
\begingroup
\renewcommand{\arraystretch}{1.2} 
\begin{table}[hbt]
\begin{center}
\caption{Number of the expected events from different sources of the background
with statistical and systematic uncertainties 
and the number of the observed events in the $ee\gamma\gamma$ and $ee\gamma$
channels after the BDT cut
}
\label{table:bkg_bdt}
\begin{tabular}{ l |  l | l }
\hline \hline
Contribution & $ee\gamma\gamma$ & $ee\gamma$ \\
\hline
$^{228}$Ac+$^{212}$Bi+$^{208}$Tl  &  7.54$\pm$0.13$\pm$0.52  & 22.83$\pm$0.26$\pm$1.60 \\
$^{214}$Bi                        &  1.65$\pm$0.04$\pm$0.38  &  6.88$\pm$0.07$\pm$1.58  \\
$^{152}$Eu+$^{154}$Eu             &  1.53$\pm$0.05$\pm$0.21  &  2.18$\pm$0.07$\pm$0.31   \\
$^{207}$Bi                        &  0.57$\pm$0.03$\pm$0.03  &  2.80$\pm$0.07$\pm$0.16   \\
$^{234m}$Pa                       &        -                 &  0.47$\pm$0.08$\pm$0.05   \\
Radon                             &  0.65$\pm$0.05$\pm$0.06  &  3.13$\pm$0.12$\pm$0.31   \\
External bkg                      &  1.70$\pm$0.43$^{+0.53}_{-0.39}$ &  3.50$\pm$0.65$^{+1.08}_{-0.80}$ \\
Neighbour foils                   &  0.12$\pm$0.01$\pm$0.03  &  0.64$\pm$0.17$\pm$0.15   \\
$^{150}$Nd $\beta\beta \to$ g.s.  &  0.12$\pm$0.01$\pm$0.01  &  3.48$\pm$0.06$\pm$0.24   \\
\hline
Total  bkg                        & 13.88$\pm$0.46$^{+0.86}_{-0.79}$ & 45.91$\pm$0.75$^{+2.56}_{-2.45}$ \\
\hline
Data                              &   53  &   85 \\
\hline
\hline
\end{tabular}
\end{center}
\end{table} 
\endgroup
\begin{figure*}[h!]
\begin{center}
\includegraphics[width=0.32\textwidth]{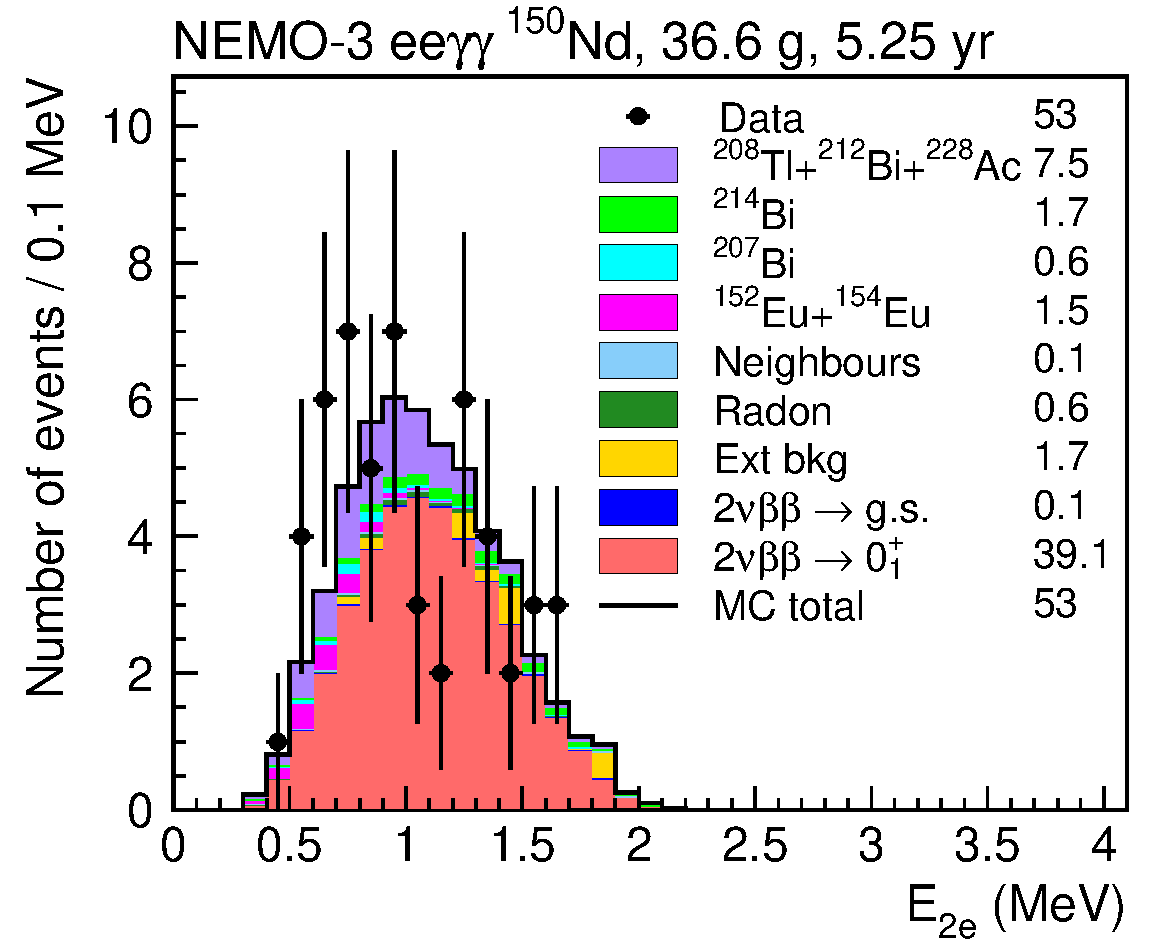}
\includegraphics[width=0.32\textwidth]{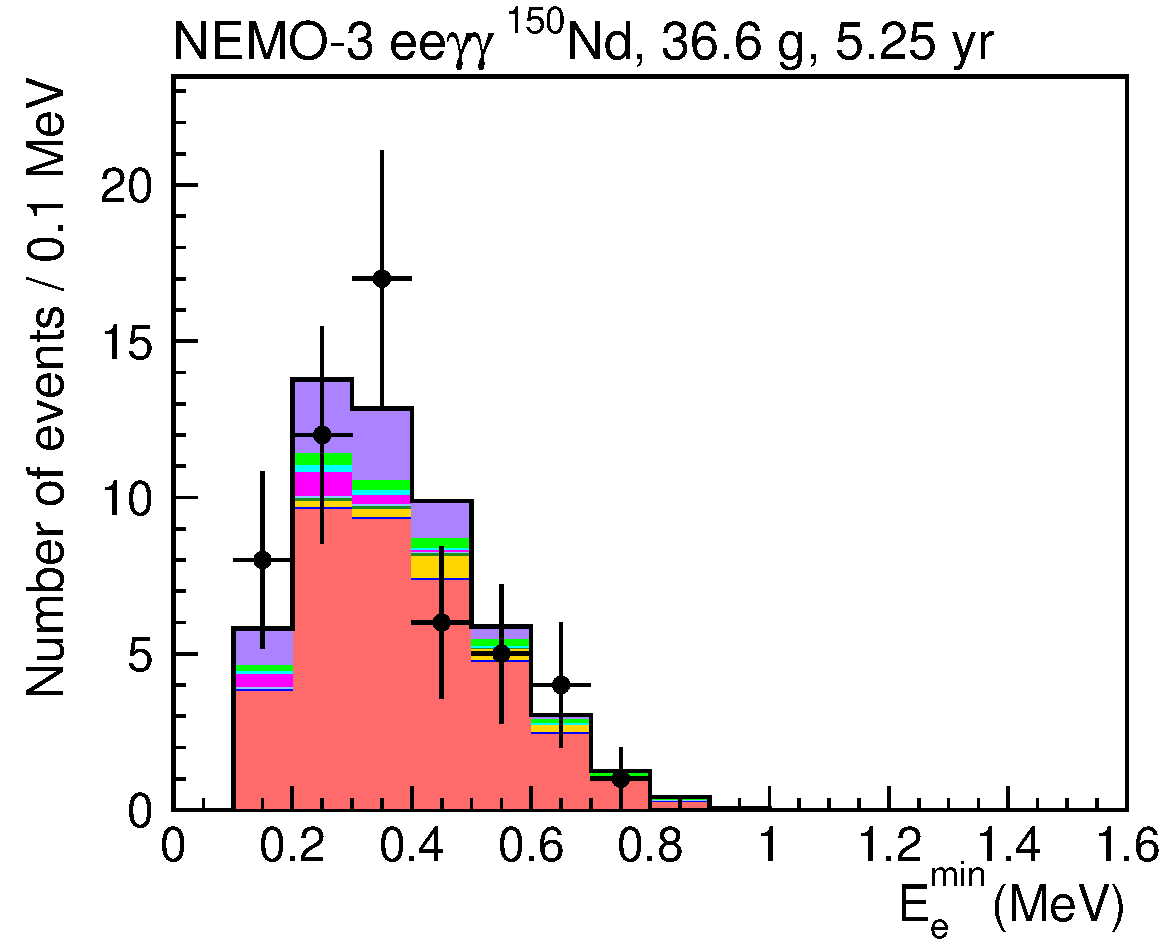}
\includegraphics[width=0.32\textwidth]{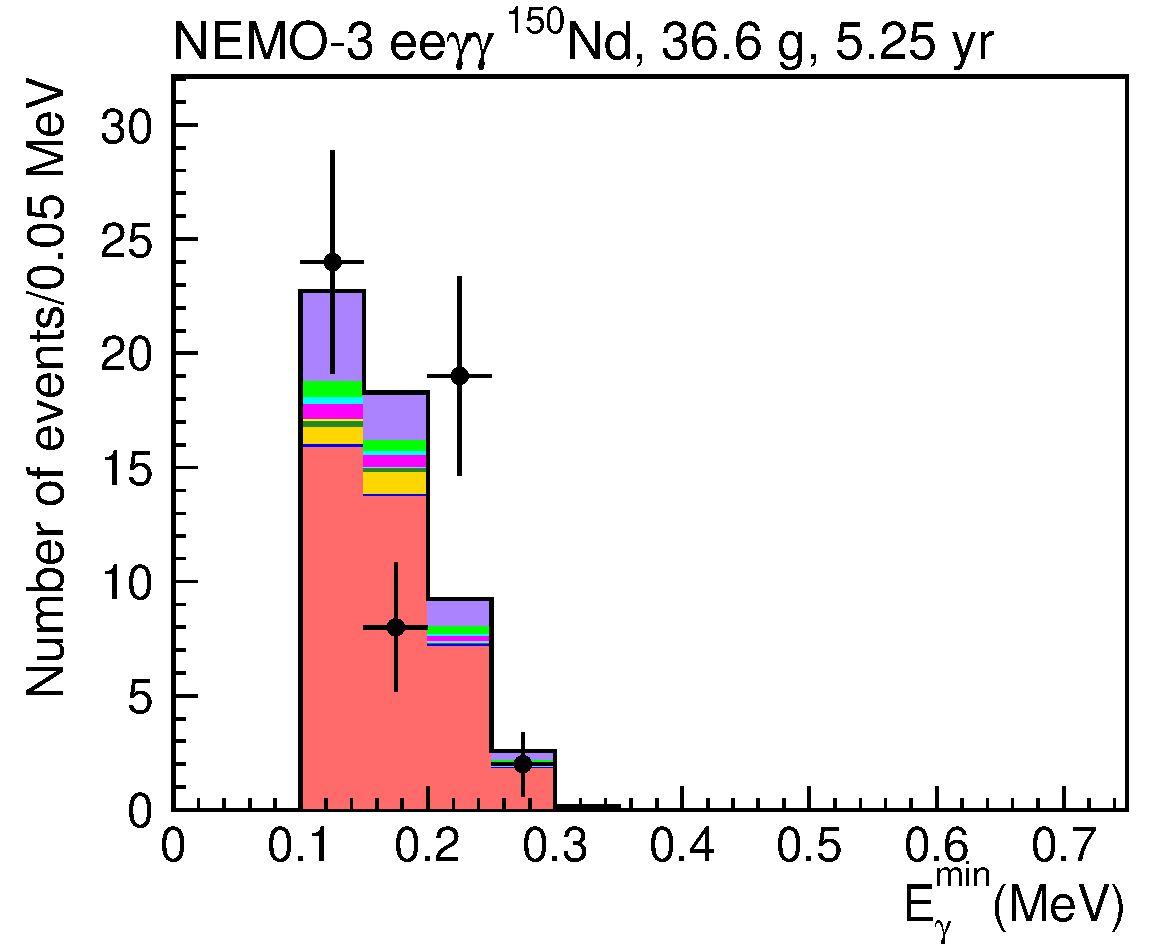}
\includegraphics[width=0.32\textwidth]{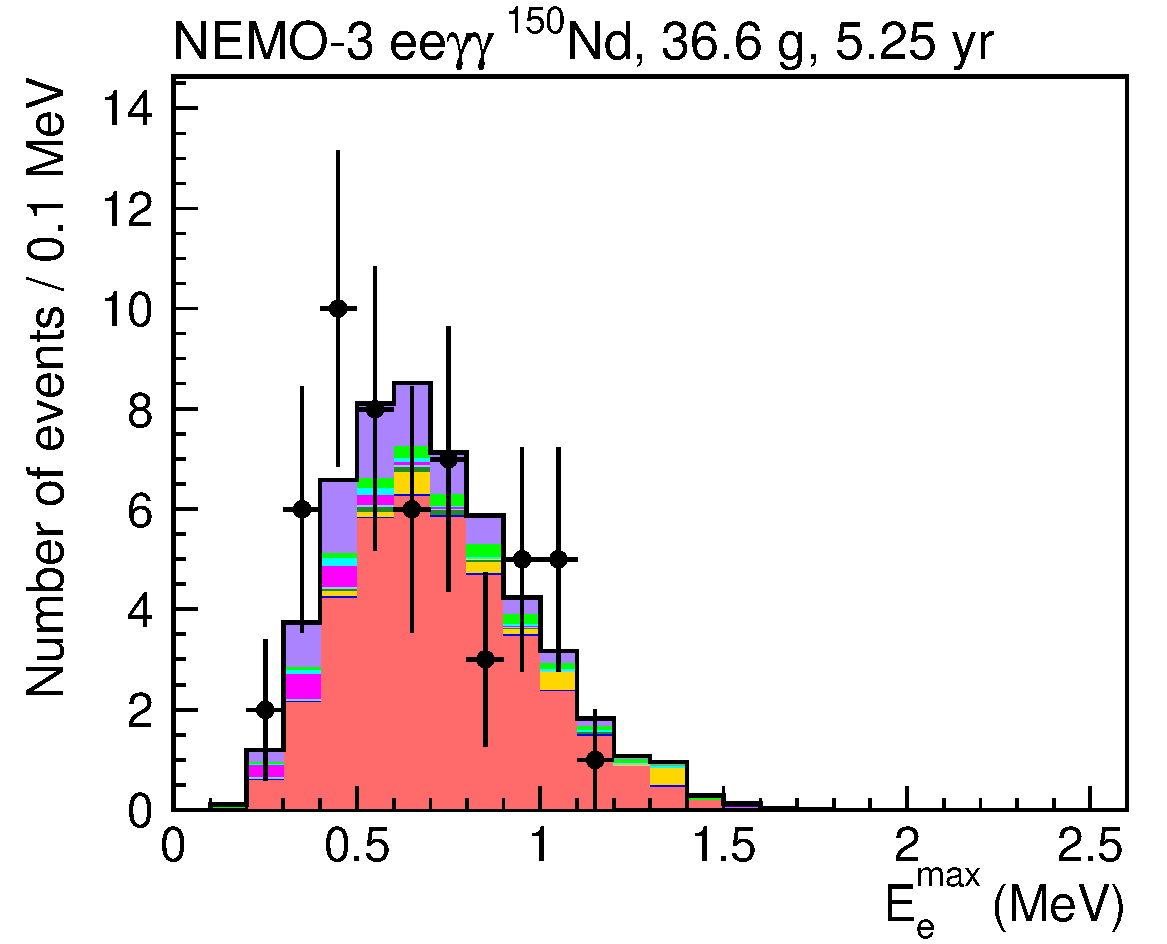}
\includegraphics[width=0.32\textwidth]{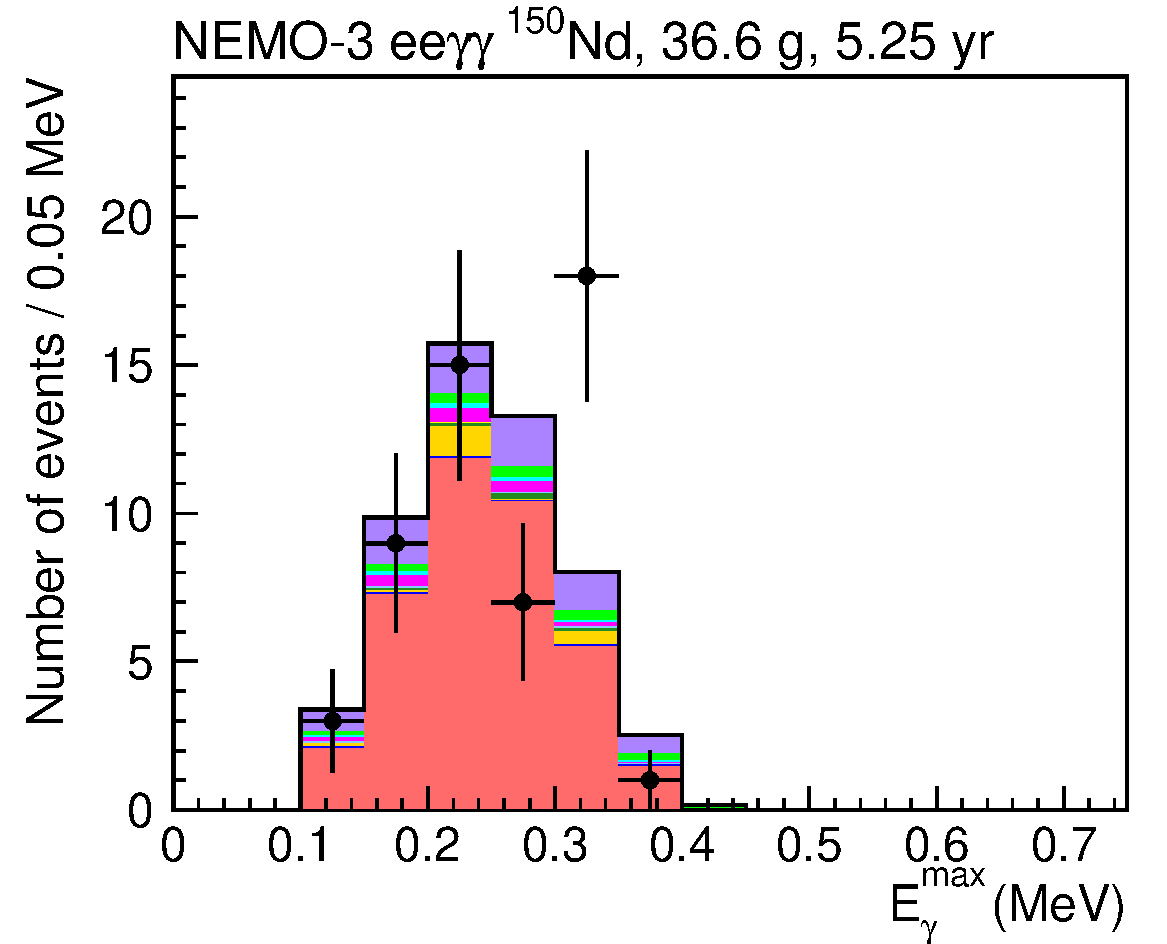}
\includegraphics[width=0.32\textwidth]{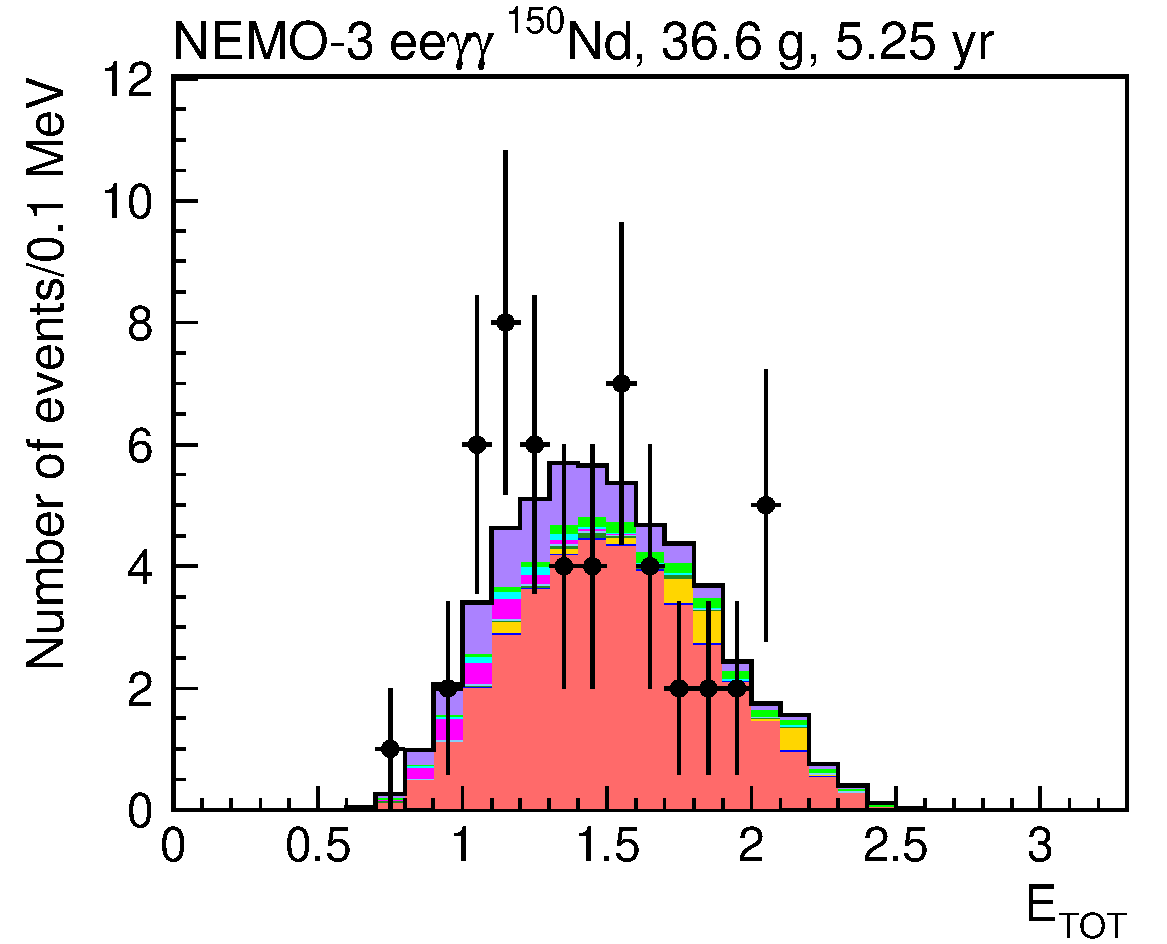}
\includegraphics[width=0.32\textwidth]{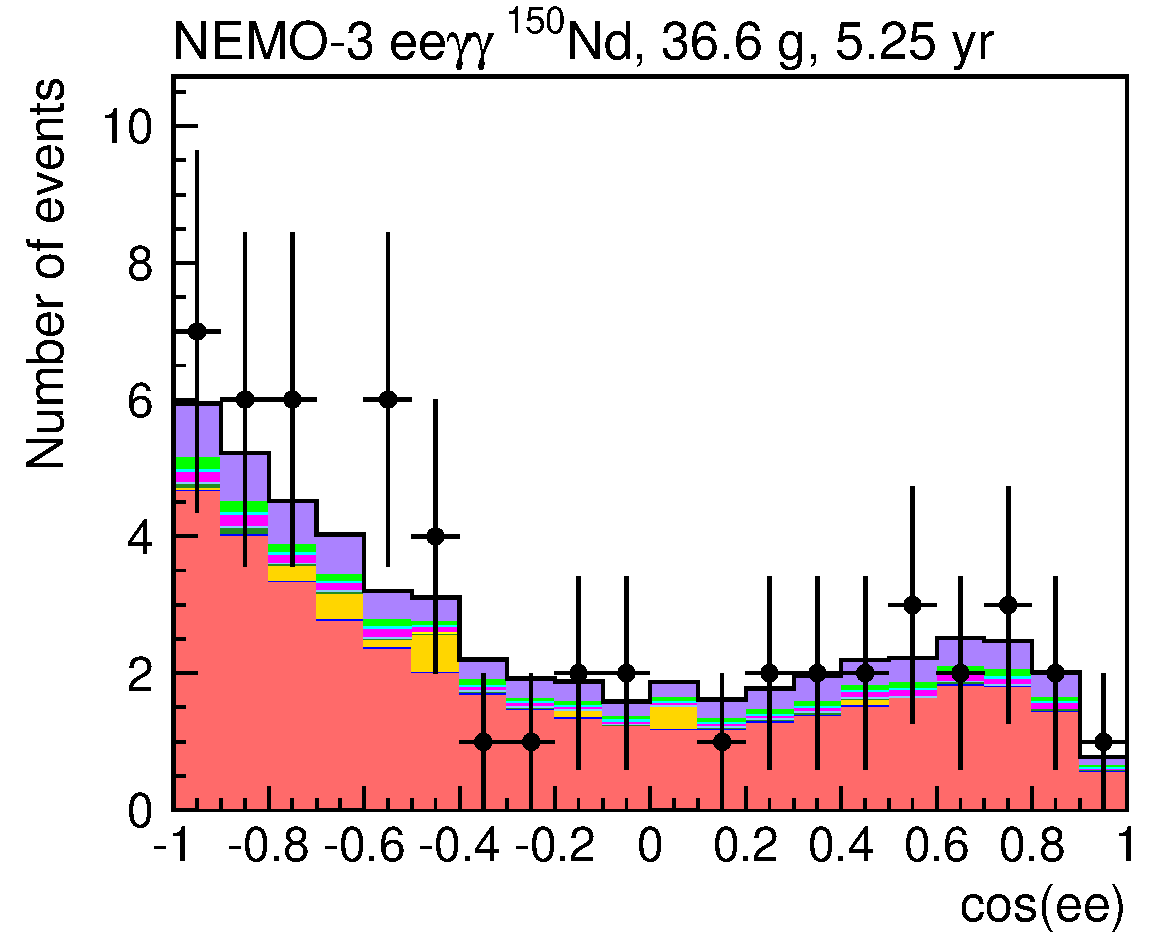}
\includegraphics[width=0.32\textwidth]{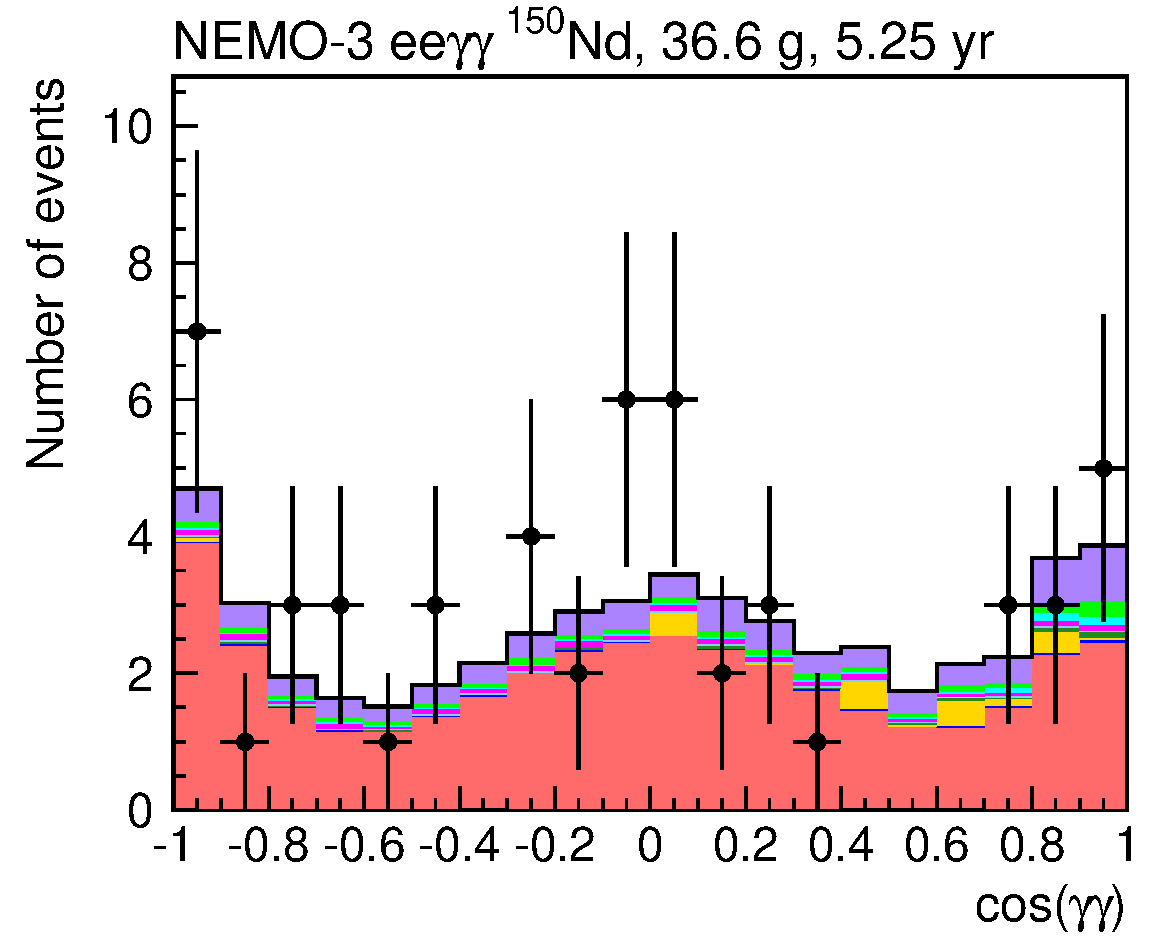}
\includegraphics[width=0.32\textwidth]{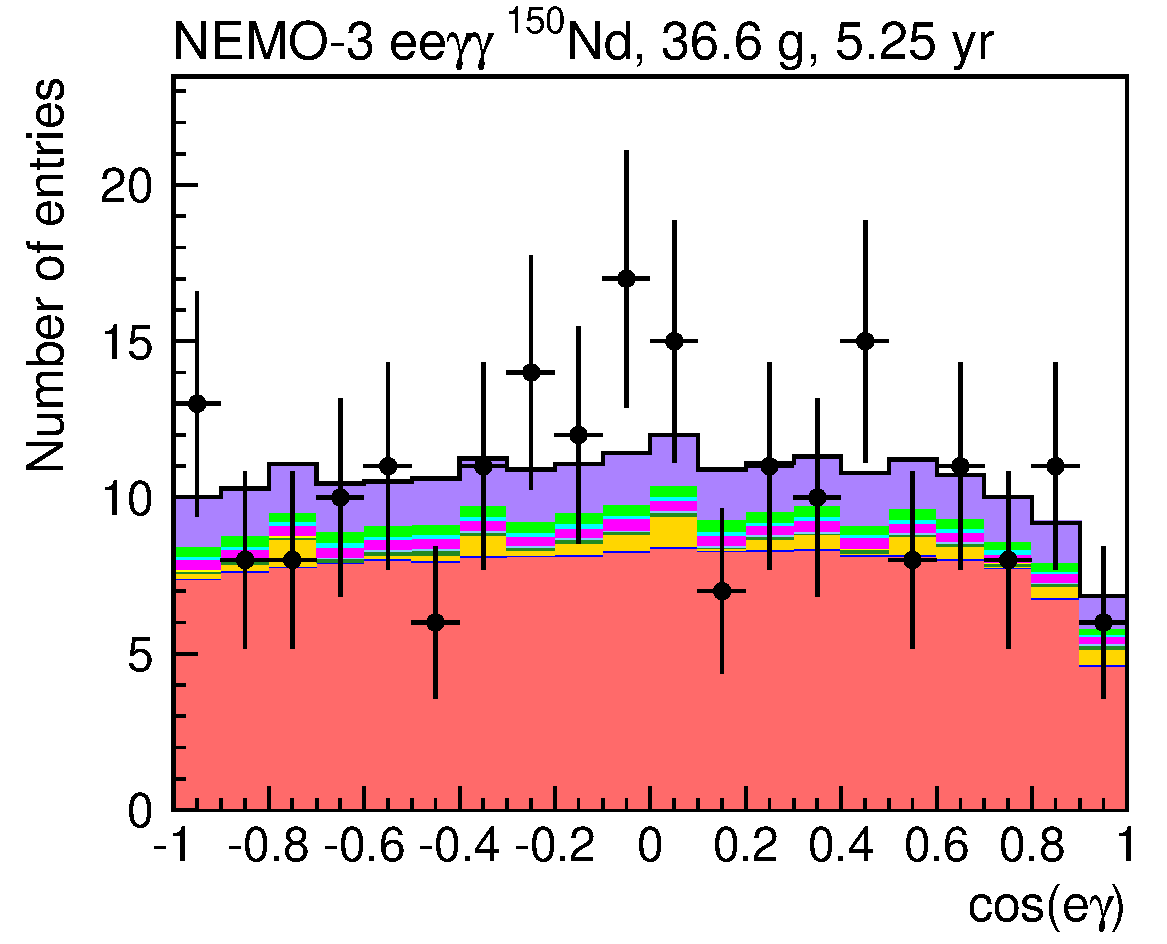}
\caption{Distributions of the two-electron two-$\gamma$ events from the $^{150}$Nd foil after the cut on BDT score: 
energy sum of two electrons $E_{2e}$, minimal electron energy $E_{e}^{\text{min}}$, minimal $\gamma$ energy $E_{\gamma}^{\text{min}}$, 
maximal electron energy $E_{e}^{\text{max}}$, maximal $\gamma$ energy $E_{\gamma}^{\text{max}}$, 
total measured energy $E_\text{TOT}$, cosine of the angle between two electrons $\cos(ee)$, between two photons 
$\cos(\gamma\gamma)$, and between electron and $\gamma$ $\cos(e\gamma)$ for all 
$e\gamma$ combinations.
 The 0$^+_1$ signal contribution is defined by performing background subtraction.}
\label{fig:eegg-0plus1b}
\end{center}
\end{figure*}
$S/B = 0.66$ and to the statistical signal significance $N{\sigma} = S/\sqrt{S+B} = 4.8$.
The 0$^+_1$ signal efficiency is $\epsilon = 0.87 \%$.
This corresponds to the following half-life estimation:
\begin{equation}
T_{1/2}^{2\nu\beta\beta}(0^{+}_{1}) = \left[ 8.18 ^{+2.18}_{-1.42}(\textrm{stat})\right]  \times10^{19}~\mbox{yr}.
\label{eqn:eegg_2nu_0plus1_prelim}
\end{equation}
In order to suppress the background and maximize the signal significance, the 
event classification employed the BDT method. Using MC of the 0$^+_1$ signal and the background, 
the BDT training is performed on the set of observables shown in Fig.~\ref{fig:eegg-prelim2},
with the total measured energy $E_{\text{TOT}}$
and the maximal $\gamma$ energy $E_{\gamma}^{\text{max}}$ being the most important variables.
After training, both the data and MC  were processed by the BDT algorithm 
which assigned a BDT score to each event  to aid discrimination of the signal from the background.
The BDT score is a continuous variable 
with lower values for more background-like events 
and higher values for more signal-like events.
The resulting BDT score distribution for the signal of the $^{150}$Nd $2\nu\beta\beta$ decay to the 0$^+_1$ excited state in the $ee\gamma\gamma$ channel is presented in Fig.~\ref{fig:bdt_eegg_2nu_0plus1}.
The vertical dashed line in this figure denotes
the optimal cut on the BDT score to maximize the signal significance. 

After rejecting the events with lower BDT score values, we are left with 53 data events and a
total expected background of 13.9 events, see Table~\ref{table:bkg_bdt}.
This requirement suppresses the background by a factor of 6.1  and reduces the signal efficiency
by a factor of 0.88 to  $\epsilon$ = 0.76\%. 
After background subtraction, 39.1 events attributed to the signal remain.
This provides the signal-to-background ratio $S/B = 2.8$ and the signal statistical significance $N{\sigma} = S/\sqrt{S+B} = 5.4$. The corresponding half-life is estimated to be
\begin{equation}
\label{eqn:eegg_2nu_0plus1}
T_{1/2}^{2\nu\beta\beta}(0^+_1) = \left[ 1.04 ^{+0.24}_{-0.16} \,\left(\mbox{stat}\right) ^{+0.12}_{-0.11}\,\left(\mbox{syst}\right 
) \right] \times10^{20}~\mbox{yr}.
\end{equation}
\begin{figure*}[h]
\begin{center}
\includegraphics[width=0.32\textwidth]{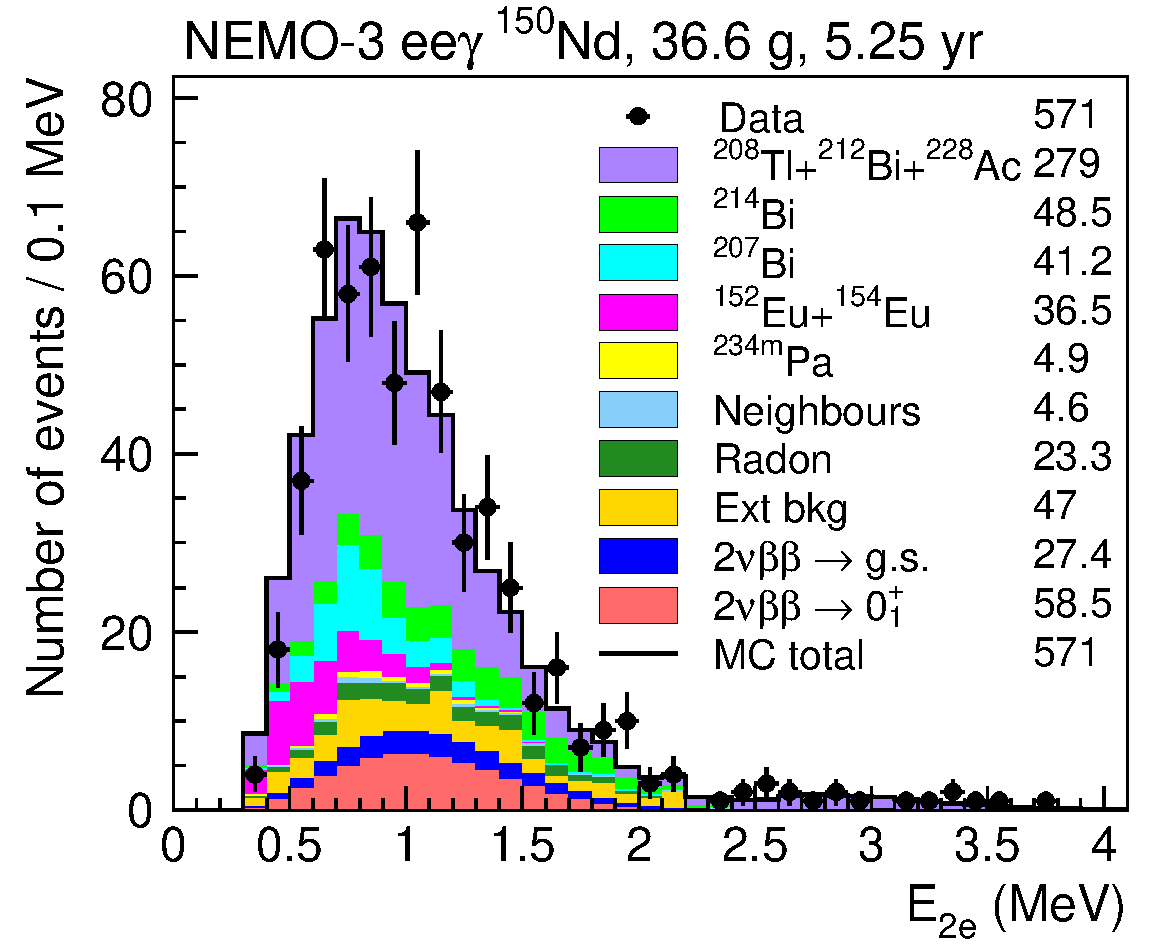}
\includegraphics[width=0.32\textwidth]{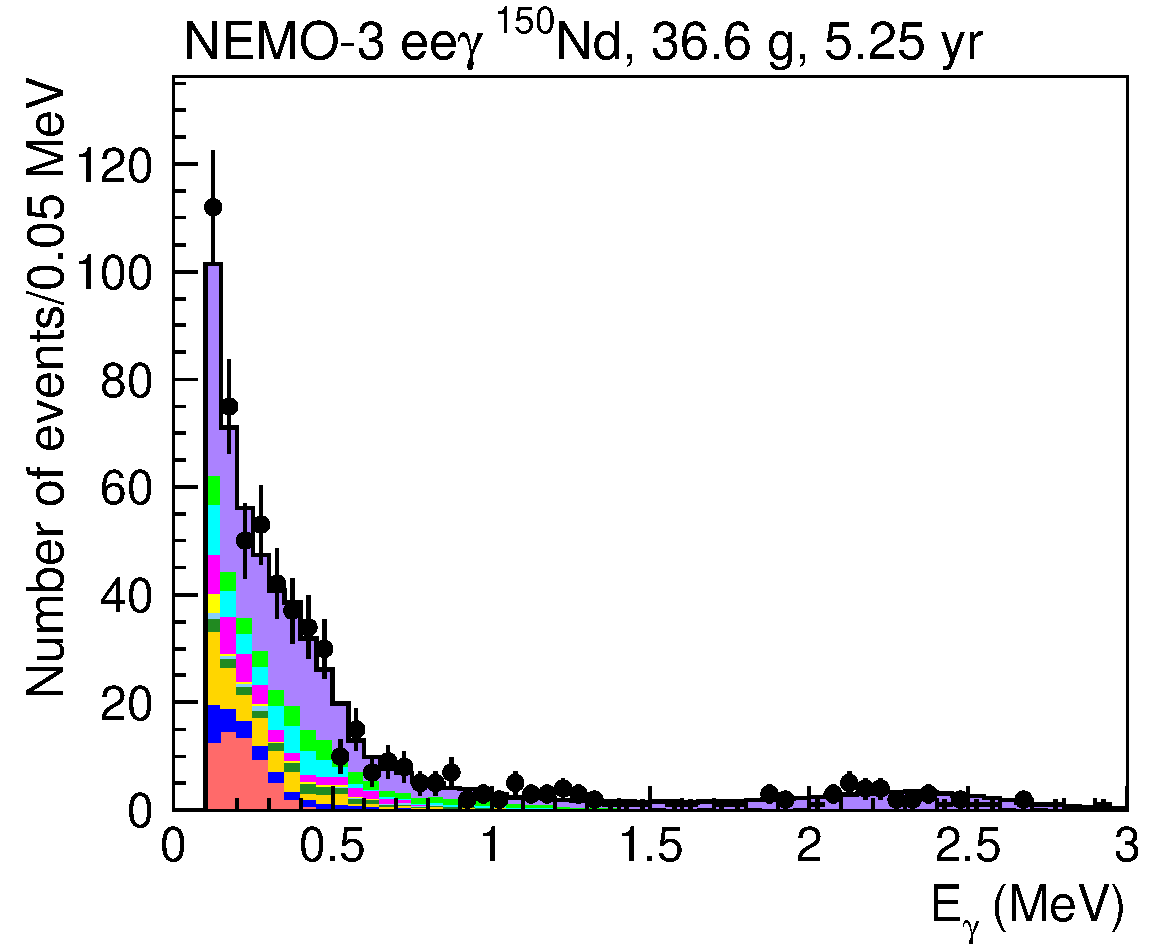}
\includegraphics[width=0.32\textwidth]{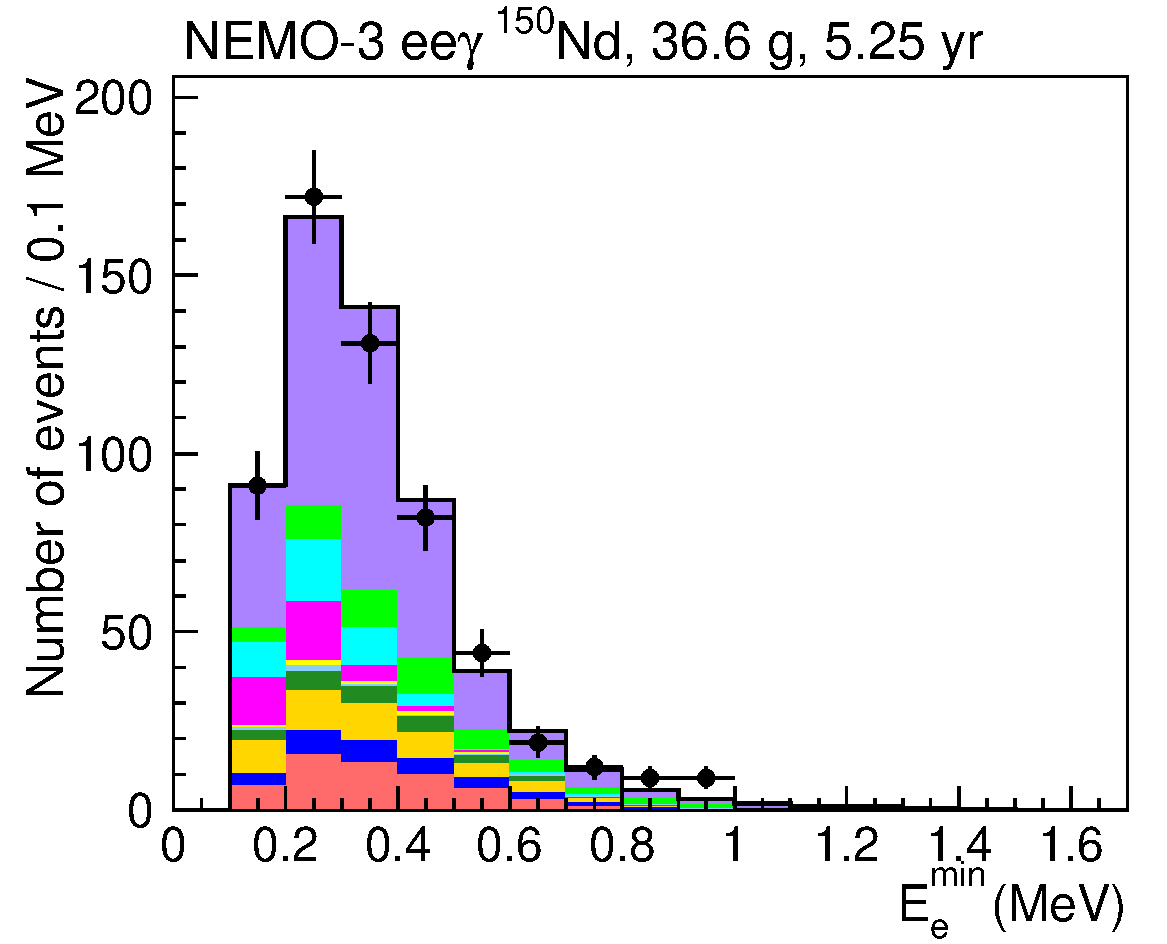}
\includegraphics[width=0.32\textwidth]{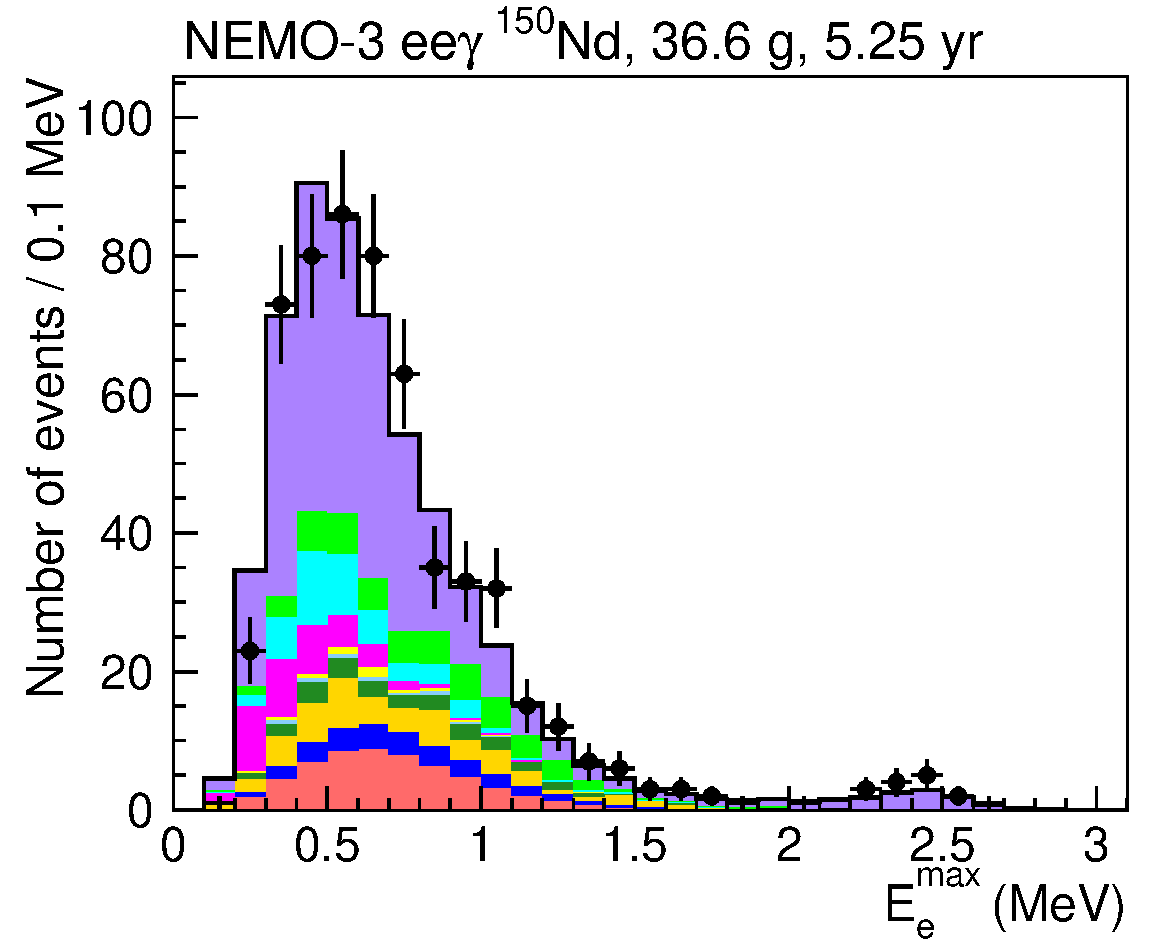}
\includegraphics[width=0.32\textwidth]{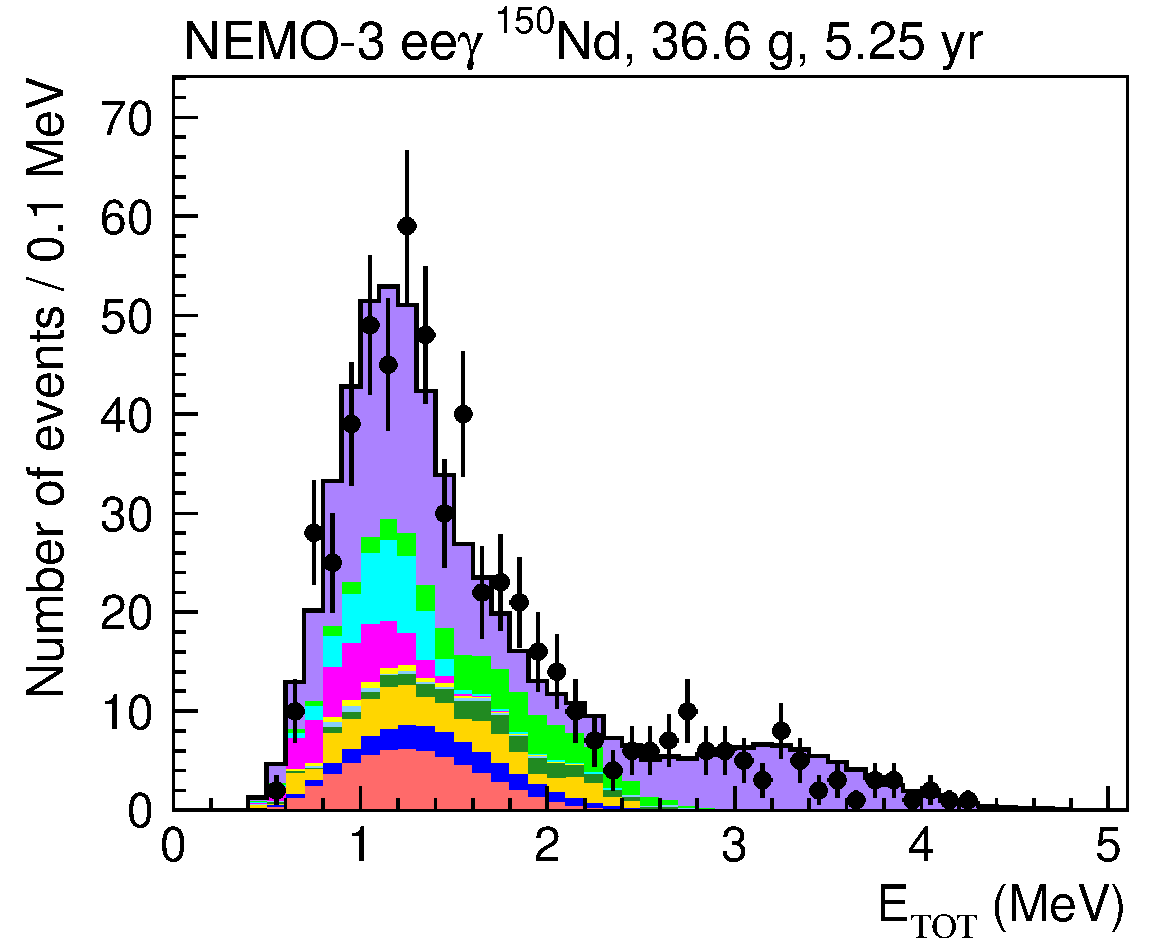}
\includegraphics[width=0.32\textwidth]{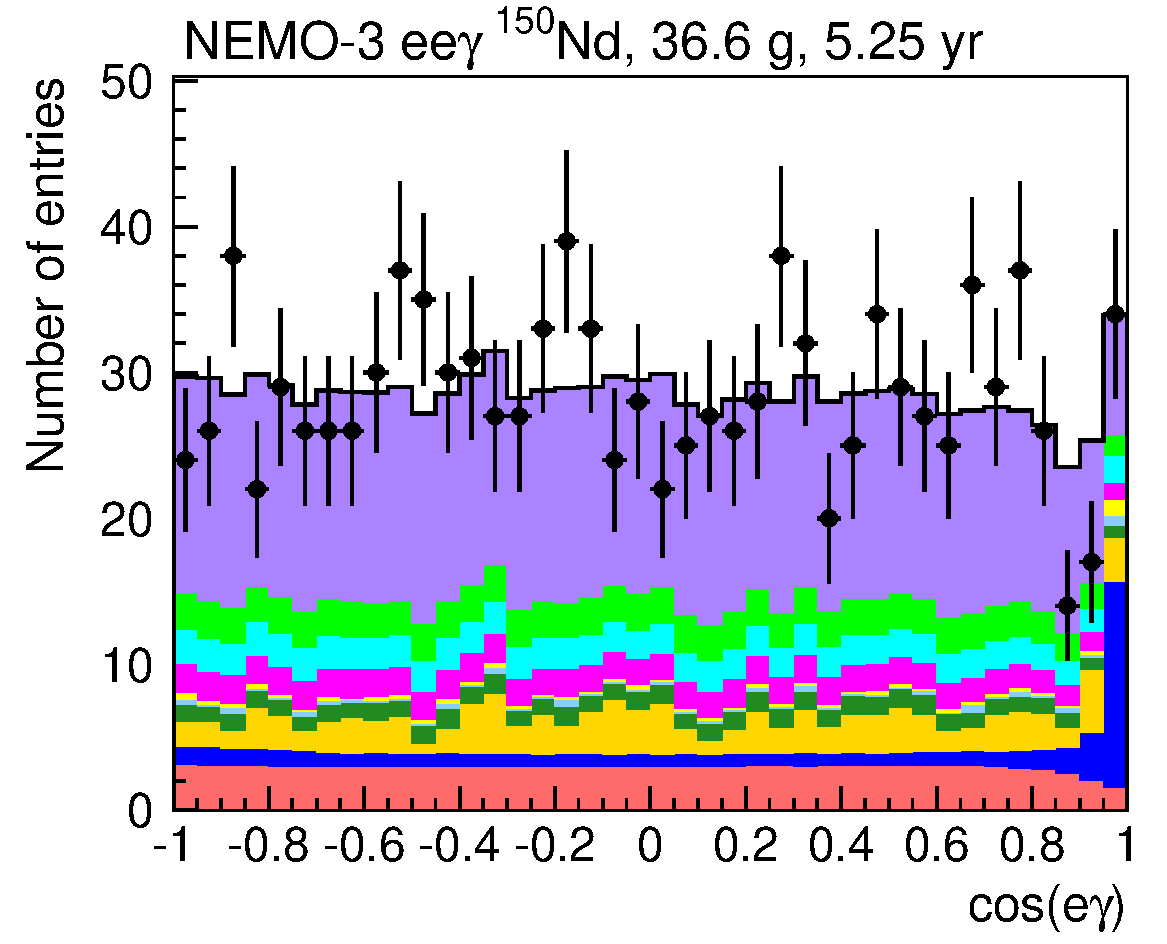}
\caption{
Distributions for the two-electron one-$\gamma$ events from the $^{150}$Nd foil
after the preliminary selection:
energy sum of two electrons $E_{2e}$, $\gamma$ energy $E_{\gamma}$, 
minimal and maximal electron energy $E_{e}^{\text{min}}$, $E_{e}^{\text{max}}$, 
total measured energy $E_\text{TOT}$, cosine of the angle between 
electron, and $\gamma$ $\cos(e\gamma)$ for both 
$e\gamma$ combinations. Data are compared to the MC prediction
with the number of $0^{+}_{1}$ signal events obtained by background subtraction.}
\label{fig:eeg-prelim}
\end{center}
\end{figure*}
This half-life value statistically agrees within 1$\sigma$ with the 
result in Eq.~\ref{eqn:eegg_2nu_0plus1_prelim} obtained after 
preliminary event selection, but is more precise.
The distributions of the measured kinematic variables 
after the cut on the BDT score are shown in 
Fig.~\ref{fig:eegg-0plus1b}.
The data show good agreement with MC for the measured quantities.
In particular, the Kolmogorov test gives a
probability of 29\% for the consistency of the data and MC
in the $\cos(\gamma\gamma$) distribution where the $\gamma\gamma$ angular correlation in MC
for the 0$^{+}$ signal is simulated according to the angular-correlation function
$W(\theta)\textrm{d}\Omega = (1-3\textrm{cos}^{2}\theta+4\textrm{cos}^4\theta)\textrm{d}\Omega$ ~\cite{Evans}  
characterizing the $0^{+} \to 2^{+} \to 0^{+}$ cascade. 
\subsubsection{Use of $\boldsymbol{ee\gamma}$ events}
The two-electron one-$\gamma$ events are also used to measure the decay to 
the 0$^+_1$ excited state since one of two emitted photons can
remain undetected. The 0$^+_1$ signal efficiency in this channel $\epsilon$ = 2.2\%
is higher than in the $ee\gamma\gamma$ channel. With $N = 571$ data events and the
total expected background of $B = 512.5$ events (see Table~\ref{table:bkg_all}), for the 0$^+_1$ signal contribution
defined by background subtraction $S = N-B = 58.5\pm23.9$ events, we obtain 
the half-life estimation
\begin{equation}
T_{1/2}^{2\nu\beta\beta}(0^{+}_{1}) = \left[ 1.98 ^{+1.37}_{-0.58} (\textrm{stat}) \right] \times10^{20}~\mbox{yr}.
\end{equation}
\begin{figure}[h]
\includegraphics[width=0.48\textwidth]{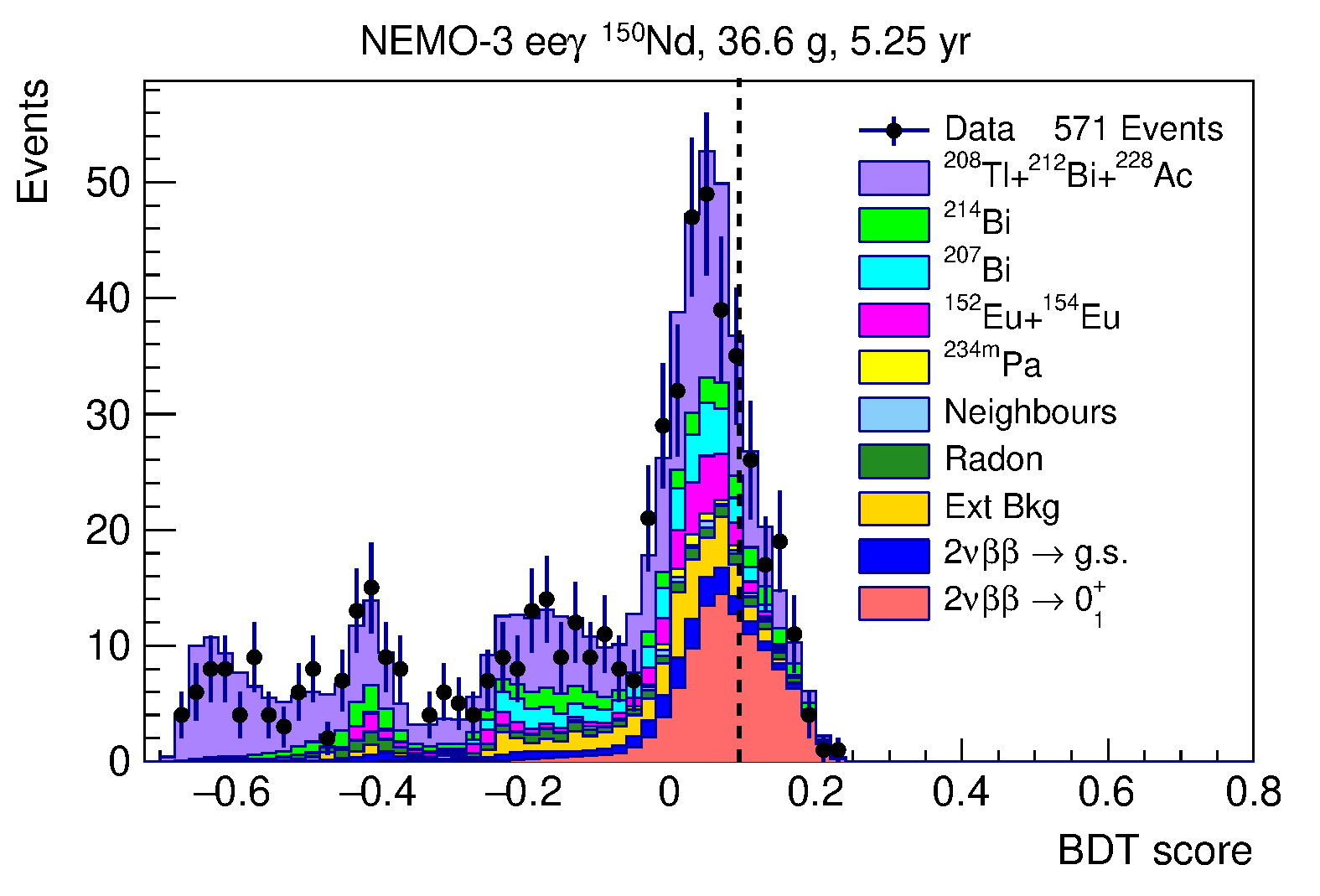}
\caption{BDT score distribution for the $2\nu\beta\beta$ decay to the 0$^+_1$ excited state
in the $ee\gamma$ channel. The vertical dashed line denotes the optimal cut position maximizing the signal significance.}
\label{fig:bdt_eeg_2nu_0plus1}
\end{figure}
\begin{figure*}[h]
\begin{center}
\includegraphics[width=0.32\textwidth]{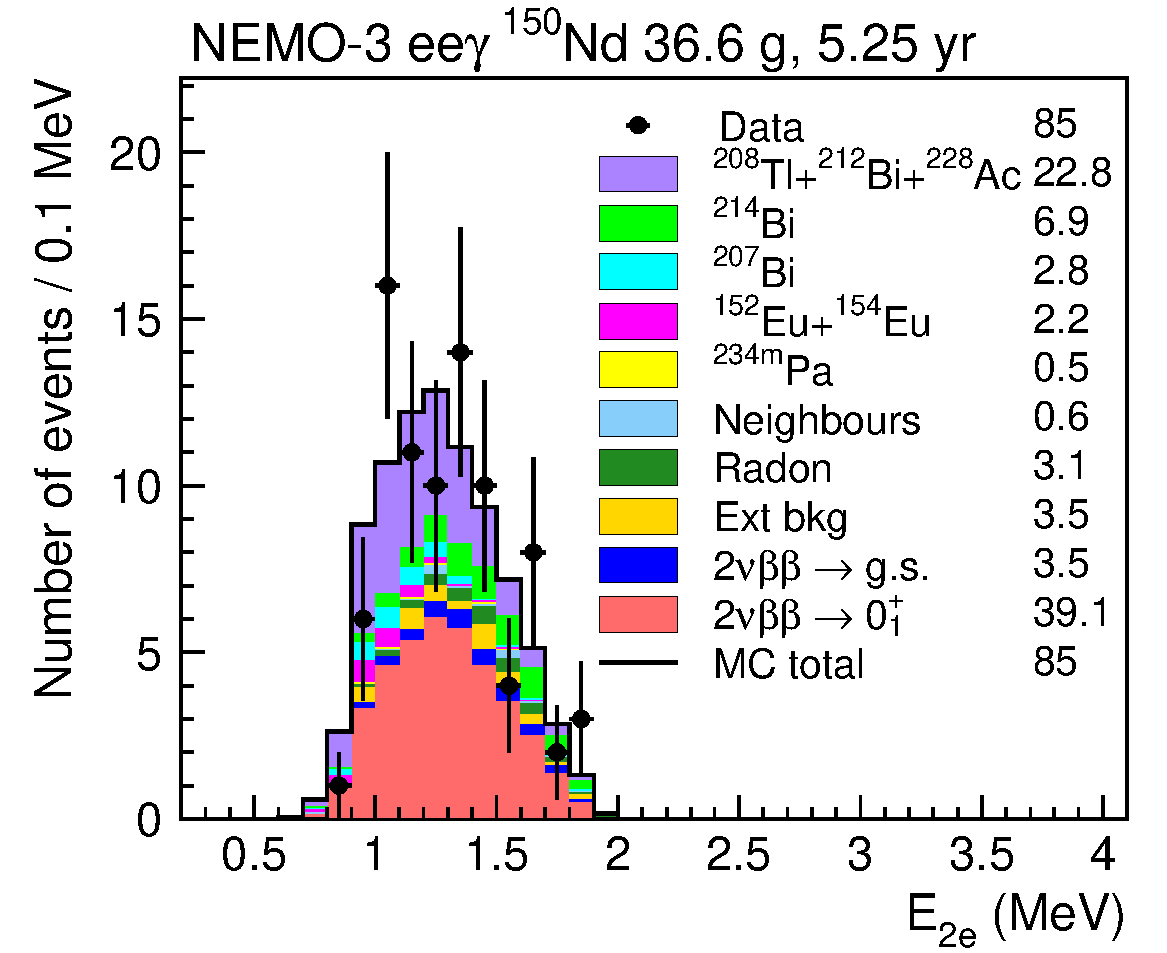}
\includegraphics[width=0.32\textwidth]{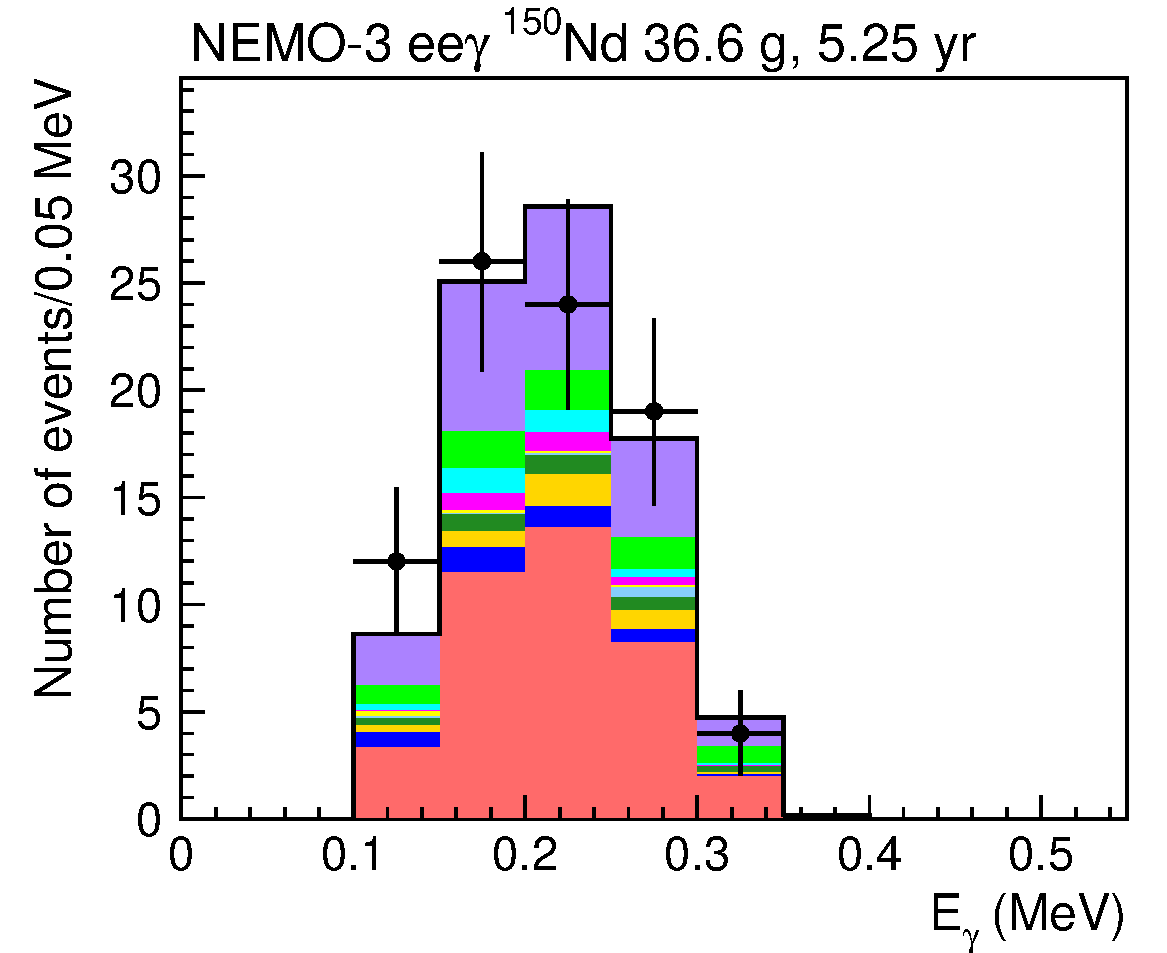}
\includegraphics[width=0.32\textwidth]{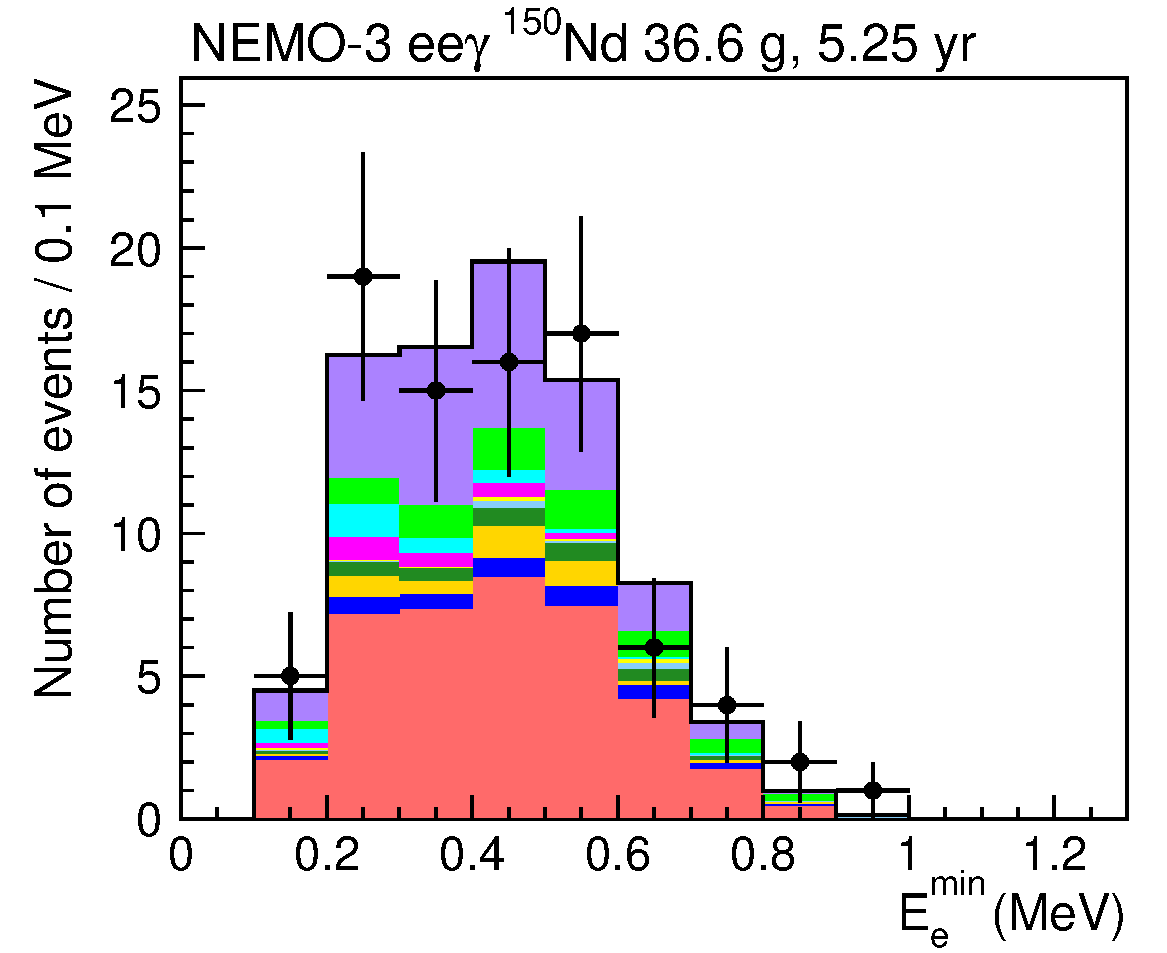}
\includegraphics[width=0.32\textwidth]{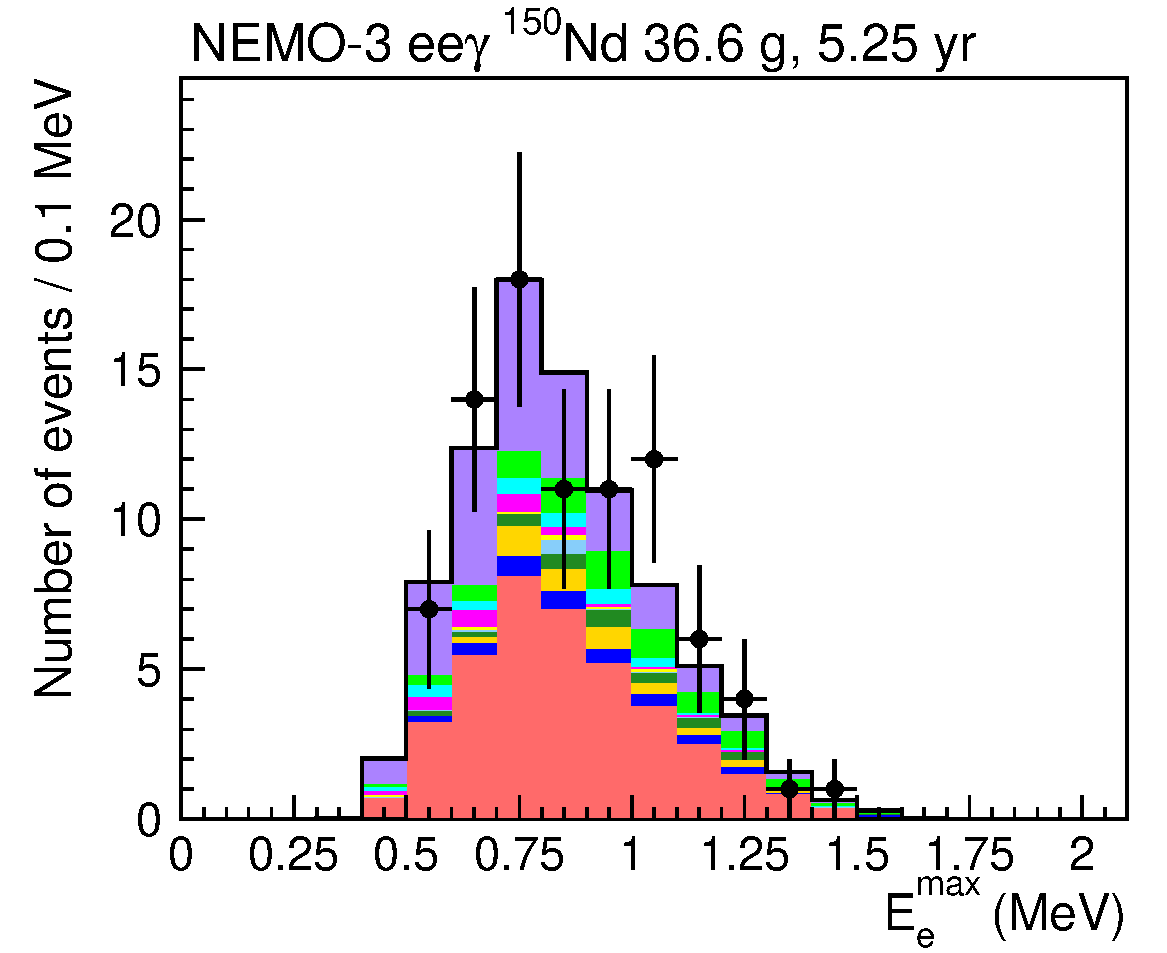}
\includegraphics[width=0.32\textwidth]{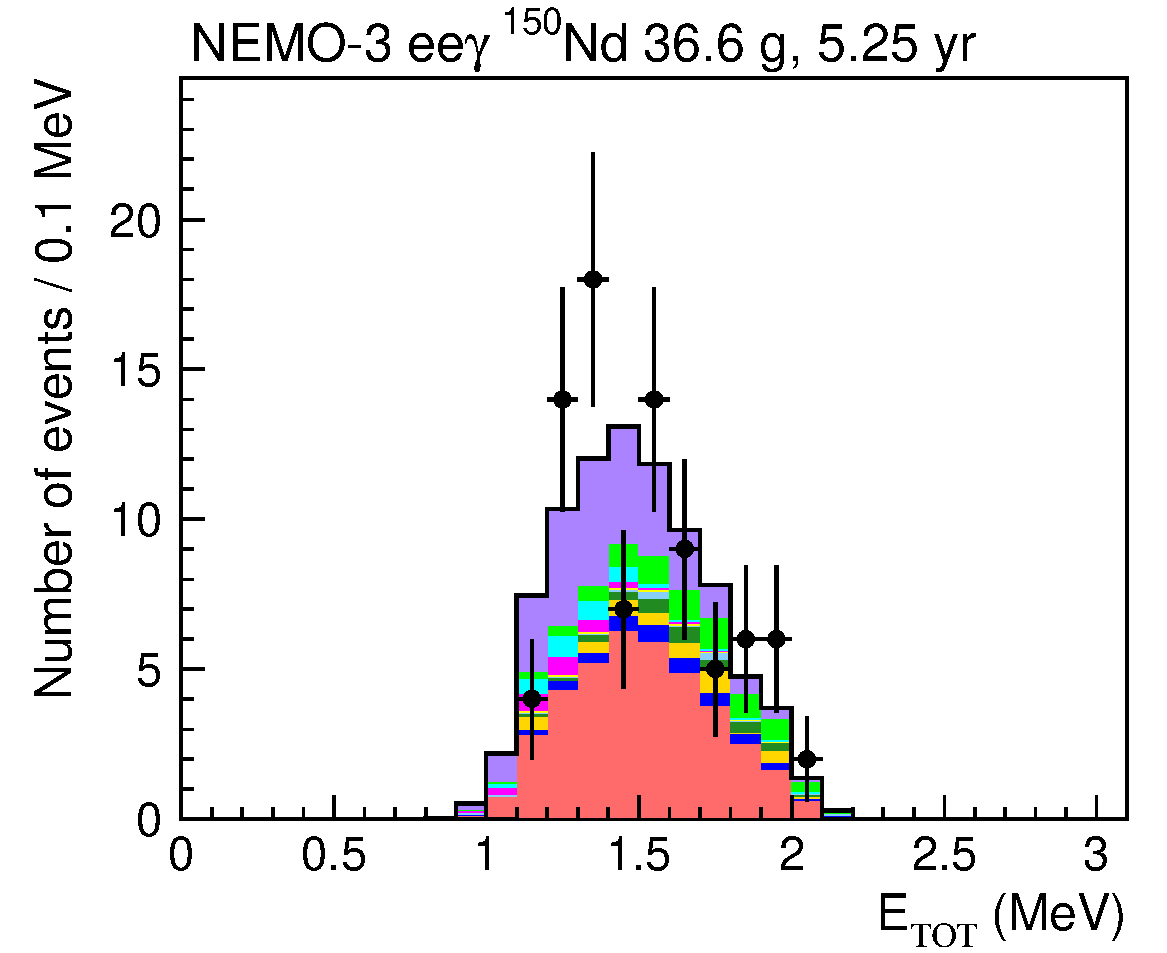}
\includegraphics[width=0.32\textwidth]{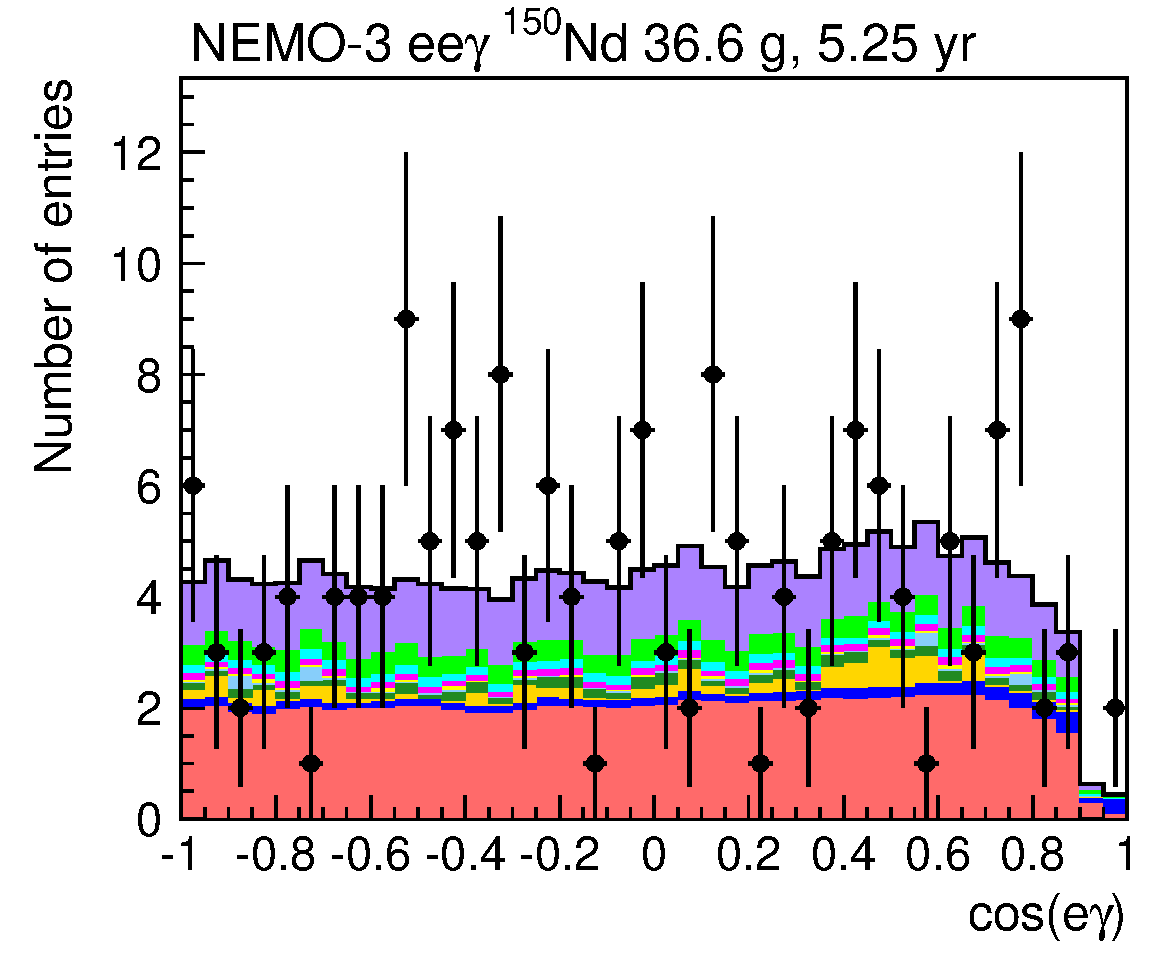}
\caption{Distributions for the two-electron one-$\gamma$ events from the $^{150}$Nd foil
after the cut on the BDT score value:
energy sum of two electrons $E_{2e}$, $\gamma$ energy $E_{\gamma}$, 
minimal and maximal electron energy $E_{e}^{\text{min}}$, $E_{e}^{\text{max}}$, 
total measured energy $E_\text{TOT}$, cosine of the angle between the
electron and $\gamma$ $\cos(e\gamma)$ for both 
$e\gamma$ combinations.
The 0$^+_1$ signal contribution is defined as the 
difference between the data and the background.}
\label{fig:eeg-0plus1}
\end{center}
\end{figure*}
This is less precise than the estimate for the  $ee\gamma\gamma$ channel due to the worse signal-to-background ratio $S/B = 0.11$ 
and the low signal statistical significance $N{\sigma} = S/\sqrt{N} = 2.5$ in this channel. 
The effect of systematic uncertainty on background rates also becomes more significant with  larger backgrounds.
Nevertheless, this estimation is statistically compatible with the measurement obtained using the $ee\gamma\gamma$ events.

The background decomposition for the selected events is presented
in Table~\ref{table:bkg_all} and the measured energy and angular 
distributions are shown in Fig.~\ref{fig:eeg-prelim}.
A possible contribution from the $\beta\beta$-decay of $^{150}$Nd to the 2$^+_1$ level is neglected.
If the $0^+_1$ contribution is normalized to the half-life value obtained
in the $ee\gamma\gamma$ channel, 
the resulting data deficit does not leave space for the
2$^+_1$ contribution. 

The 
photon energy $E_{\gamma}$ and the total measured energy $E_\text{TOT}$ are the most important variables in the set of observables used for BDT training in this channel.
The BDT score distribution obtained after the event classification is shown in Fig.~\ref{fig:bdt_eeg_2nu_0plus1}. The position of the optimal BDT cut to
maximize the 0$^+_1$ signal significance is marked by a vertical dashed line.
After the BDT cut,  85 data events remain, with  $B = 45.9$ expected background events (see Table~\ref{table:bkg_bdt}).
Subtracting the background leaves $S =39.1$ events attributed to the 0$^+_1$ signal.
The resulting signal-to-background ratio is $S/B = 0.85$, with a statistical signal significance of
$N{\sigma} = S/\sqrt{S+B} = 4.2$. 
The distributions of the measured kinematic variables for these events 
are shown in Fig.~\ref{fig:eeg-0plus1}. The signal efficiency after the
BDT cut is $\epsilon = 0.88\%$ and the half-life estimate is
\begin{equation}
\label{eqn:eeg_2nu_0plus1}
T_{1/2}^{2\nu\beta\beta}(0^+_1) = \left[ 1.21 ^{+0.37}_{-0.23} \,\left(\mbox{stat}\right) ^{+0.26}_{-0.20}\,\left(\mbox{syst}\right) \right] \times10^{20}~\mbox{yr}.
\end{equation}   
\subsubsection{Systematic uncertainties}
Several sources of systematic uncertainty were investigated.
The largest contribution to the systematic uncertainty of the measured 
decay rate comes from the uncertainty on the 
event selection efficiency.
This uncertainty is estimated by measuring the calibrated $^{232}$U source activities. 
These activities, measured 
by the NEMO-3 detector for $ee\gamma\gamma$ and $ee\gamma$ events, 
are found to be in agreement with 
true values within 7.8\%.
This uncertainty, taken into account both for signal and background events,
leads to the decay rate uncertainty of (+11.5, --9.8)\% in the $ee\gamma\gamma$ 
and (+18.4, --15.7)\%  in the $ee\gamma$ channel.

The systematic uncertainty on the number of background events
(see Tables~\ref{table:bkg_all}, \ref{table:bkg_bdt}) was calculated from 
the systematic uncertainties on the individual background component activities
estimated in \cite{Arnold:2016nd150}. The background uncertainty contributes to the decay rate uncertainty of (+2.3, --2.5)\% 
in  the $ee\gamma\gamma$  
and (+6.6, --6.8)\% in the $ee\gamma$ channel. 

The effect of the limited accuracy in simulation of ionization energy loss 
and of bremsstrahlung in the foil on the measured decay rate was studied.
This was done by generating additional MC data samples varying the
relevant parameters within their expected uncertainty. 
The decay rate uncertainty due to
ionization energy loss was evaluated to be $\pm$1.6\%  in the $ee\gamma\gamma$
and $\pm$2.8\% in  the $ee\gamma$ channel. The uncertainty due to bremsstrahlung
is $\pm$1.2\% in the $ee\gamma\gamma$ and $\pm$4.4\%  in the $ee\gamma$ channel.

The effect of energy calibration uncertainty was studied by altering
measured energies according to the energy scale uncertainty; it yields a 
systematic uncertainty of $\pm$1\% on the decay rate measurement in the $ee\gamma\gamma$
and  $\pm$1.6\%  in the $ee\gamma$ channel.

Finally, a $\pm$0.5\% uncertainty on the mass of $^{150}$Nd  
translates into the same uncertainty on the measured decay rate.

All these contributions are summarized in Table~\ref{table:syst_0plus1},
with the total uncertainty calculated by summing the individual contributions
in quadrature.

\begingroup
\renewcommand{\arraystretch}{1.4} 
\begin{table}[hbt]
\begin{center}
\caption{Sources of systematic uncertainty on the rate $\Gamma$ of $2\nu\beta\beta$ decay to the 0$^+_1$ excited state
measured with $ee\gamma\gamma$ and $ee\gamma$ events. }
\label{table:syst_0plus1}
\begin{tabular}{ p{2.8 cm} | c | c | c }
\hline \hline
  & \multicolumn{3}{ c }{$\Delta\Gamma\,[\%]$} \\
\hline
Contribution &  $ ee\gamma\gamma$ &  $ ee\gamma$& mean\\
\hline
 Efficiency               & $^{+11.5}_{-9.8}$ & $^{+18.4}_{-15.7}$ & $^{+14.4}_{-12.3}$ \\
 Total background         & $^{+2.3}_{-2.5}$  & $^{+6.6}_{-6.8}$   & $^{+4.1}_{-4.3}$ \\
 Energy loss in foil      & $\pm 1.6$     & $\pm 2.7$ & $\pm 2.1$ \\
 Energy calibration       & $\pm 1  $     & $\pm 1.6  $ & $\pm 1.3$ \\
 Bremsstrahlung modelling & $\pm 1.2$     & $\pm 4.4$ & $\pm 2.5$ \\
 Mass of $^{150}$Nd       & $\pm 0.5$     & $\pm 0.5$ & $\pm 0.5$ \\
\hline
 Total                    & $^{+11.9}_{-10.4}$ & $^{+20.3}_{-18}$ & $^{+15.4}_{-13.5}$ \\
\hline
\hline
\end{tabular}
\end{center}
\end{table} 
\endgroup
\subsubsection{Mean half-life from $\boldsymbol{ee\gamma\gamma}$ and $\boldsymbol{ee\gamma}$ channels}
The individual half-life estimates from the $ee\gamma\gamma$ and $ee\gamma$ channels in
Eq.~\ref{eqn:eegg_2nu_0plus1} and Eq.~\ref{eqn:eeg_2nu_0plus1} are in  good agreement.
The mean value of the two measurements was calculated using their statistical weights:
\begin{equation}
\label{eqn:mean_2nu_0plus1}
T_{1/2}^{2\nu\beta\beta}(0^+_1) = \left[ 1.11 ^{+0.19}_{-0.14} \,\left(\mbox{stat}\right) ^{+0.17}_{-0.15}\,\left(\mbox{syst}\right) \right] \times10^{20}~\mbox{yr}.
\end{equation}
The systematic uncertainty of the mean value was obtained by calculating the
mean of the two measurements coherently increased/decreased by their 
individual systematic uncertainties.
The mean value corresponds to the signal-to-background ratio $S/B = 1.3$ and the
statistical signal significance $N\sigma_{\textrm{stat}}$ = 6.8.
Using the total error determined by summing the statistical and systematic errors in quadrature,
we obtain the 0$^+_1$ signal significance $N\sigma_\textrm{tot}$ = 5.
This half-life value is compared with the results of the previous measurements in Table~\ref{table:compare_t12}.
There is a good overall agreement between the results. The value obtained in this work is 
most precise.
\begingroup
\renewcommand{\arraystretch}{1.4} 
\begin{table}[hbt]
\begin{center}
\caption{Comparison of our results for  $^{150}$Nd $\beta\beta$ decay to the $0^{+}_{1}$ and $2^{+}_{1}$ excited states
of $^{150}$Sm with the previous results. Limits are given at 90\% C.L.}
\label{table:compare_t12}
\begin{tabular}{ c  l  l }
\hline \hline
Decay & $T_{1/2}$, 10$^{20}$yr & Reference \\
\hline   
2$\nu\beta\beta \to$ 0$^+_1$   &   $ 1.11 ^{+0.19}_{-0.14}\,(\mbox{stat}) ^{+0.17}_{-0.15}\,(\mbox{syst}) $ & This work \\
2$\nu\beta\beta \to$ 0$^+_1$  & $1.33 ^{+0.36}_{-0.23}\,(\mbox{stat})^{+0.27}_{-0.13}\,(\mbox{syst})$ & \cite{PhysRevC.79.045501}  \\     
2$\nu\beta\beta \to$ 0$^+_1$  & $1.07 ^{+0.45}_{-0.25}\,(\mbox{stat})\pm{0.07}\,(\mbox{syst})$ & \cite{PhysRevC.90.055501}  \\     
2$\nu\beta\beta \to$ 0$^+_1$  & $0.97 ^{+0.29}_{-0.19}\,(\mbox{stat})\pm{0.15}\,(\mbox{syst})$  & \cite{Polischuk}  \\ 
\hline
0$\nu\beta\beta \to$ 0$^+_1$   & $ > 136 $            & This work \\
0$\nu\beta\beta \to$ 0$^+_1$   & $ > 2.4 $            & \cite{Nasim} \\
\hline  
2$\nu\beta\beta \to$ 2$^+_1$   & $ > 2.4 $            & This work \\
2$\nu\beta\beta \to$ 2$^+_1$   & $ > 2.2 $           &  \cite{PhysRevC.79.045501}\\  
\hline  
0$\nu\beta\beta \to$ 2$^+_1$       & $ > 126$            & This work \\  
0$\nu\beta\beta \to$ 2$^+_1$       & $ > 24$            & \cite{Nasim} \\ 
\hline \hline
\end{tabular}
\end{center}
\end{table} 
\endgroup

This  half-life value may be used to extract the experimental value 
of the corresponding NME according to Eq.~\ref{eq:one}. 
Using the phase space factor value $G = 4.116 \times 10^{-18}~\mbox{yr}^{-1}$ ~\cite{Stoica} and 
$g_A$ = 1.2756 ~\cite{Zyla:2020zbs}, 
one obtains the NME value (scaled by the electron rest mass) for the 2$\nu\beta\beta$  transition to the
0$^+_1$ excited state:
\begin{equation}
|M^{2\nu}(0^+_1)| = 0.0288^{+0.0032}_{-0.0028}  ~ ~ .
\end{equation}
One can compare this value with the NME value for the 2$\nu\beta\beta$  transition 
to the ground state of $^{150}$Sm:
\begin{equation}
|M^{2\nu}(0^+_{\textsf{g.s.}})| = 0.0338^{+0.0013}_{-0.0011} ~ ~ ,
\end{equation}
obtained using 
 the half-life of Eq.~\ref{eq:three} 
from~\cite{PhysRevC.79.045501} and $G = 3.540 \times 10^{-17} yr^{-1}$ from~\cite{Stoica}). 
The relative difference between these two NMEs is  $(15^{+10}_{-9})$\%.
One can see that the values of these matrix elements are very close (in the first approximation they are the same) and that $M^{2\nu}(0^+_{\textsf{g.s.}})$ 
is $\sim$15\% greater than $M^{2\nu}(0^+_{1})$. 
This is consistent with the conclusions in~\cite{BarHubHub04},\cite{nemo3-mo100-excited}. 
The same conclusions were drawn
for  $^{100}$Mo (see, for example,\cite{nemo3-mo100-excited}). Thus, it looks like a certain regularity. 
It seems important
and interesting to understand this regularity and to give it some theoretical explanation. Finally, 
this can help to clarify the situation with the calculations of nuclear matrix elements.
\subsection{Search for $\boldsymbol{2\nu\beta\beta}$ decay to $\boldsymbol{2^+_1}$ excited state}
As stated above, there is no excess in the data above the expected background 
that could be attributed to the $2\nu\beta\beta$ decay to the 
2$^+_1$  excited state. A limit on this transition was set using the $ee\gamma$
events. A BDT classification was performed for the 2$^+_1$ signal against backgrounds, including all the contributions listed in Table~\ref{table:bkg_all} plus the $2\nu\beta\beta$ decay to the 0$^+_1$ excited state normalized according to Eq.~\ref{eqn:mean_2nu_0plus1}.
\begin{figure}[h]
\includegraphics[width=0.48\textwidth]{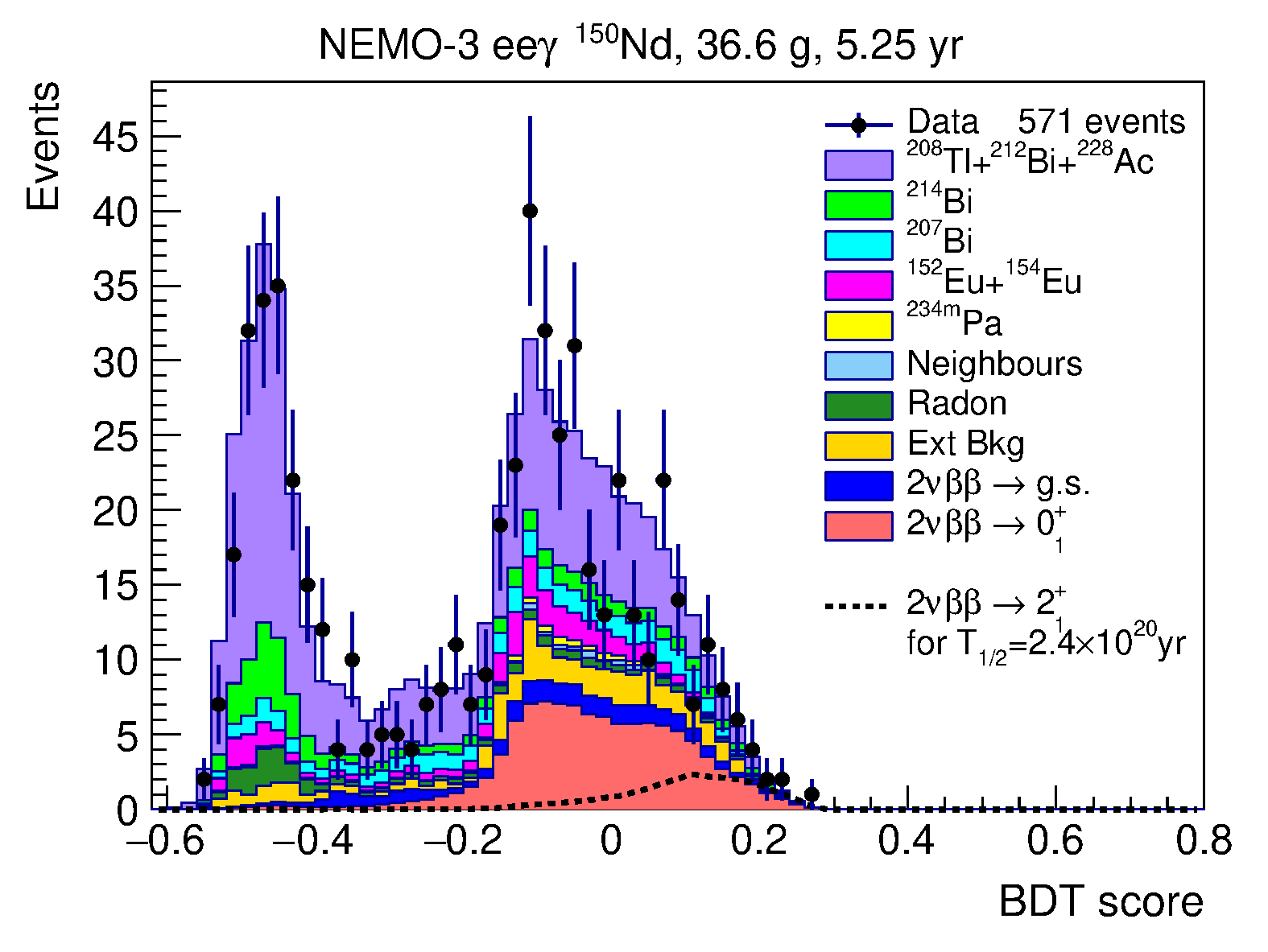}
\caption{BDT score distribution for the $ee\gamma$ events used to set the limit
on the $^{150}$Nd $2\nu\beta\beta$ decay to the 2$^+_1$ excited state
of $^{150}$Sm}
\label{fig:bdt_eeg_2nu}
\end{figure}
The BDT score distribution in Fig.~\ref{fig:bdt_eeg_2nu} was used for calculating the limit. 
Hereafter, the systematic uncertainties discussed above 
are taken into account.
Systematic uncertainties are treated as uncertainties on the 
expected numbers of events and are folded
into the signal and background expectations 
via the Gaussian distribution. 
The systematic uncertainties considered for each of the
background component normalizations are given in Table~\ref{table:bkg_all}.
The uncertainty on the normalization of the $^{150}$Nd 2$\nu\beta\beta$ decay 
to the 0$^+_1$ excited state is constrained by 
the statistical uncertainty of the measurement given in
Eq.~\ref{eqn:mean_2nu_0plus1}.
The obtained result 
\begin{equation}
T^{2\nu\beta\beta}_{1/2}(2^+_1) > 2.42 \times 10^{20}~\mbox{yr at 90\% C.L.}
\end{equation}
is slightly more restrictive than the limit set in~\cite{PhysRevC.79.045501}.
\begin{figure}[h]
\includegraphics[width=0.48\textwidth]{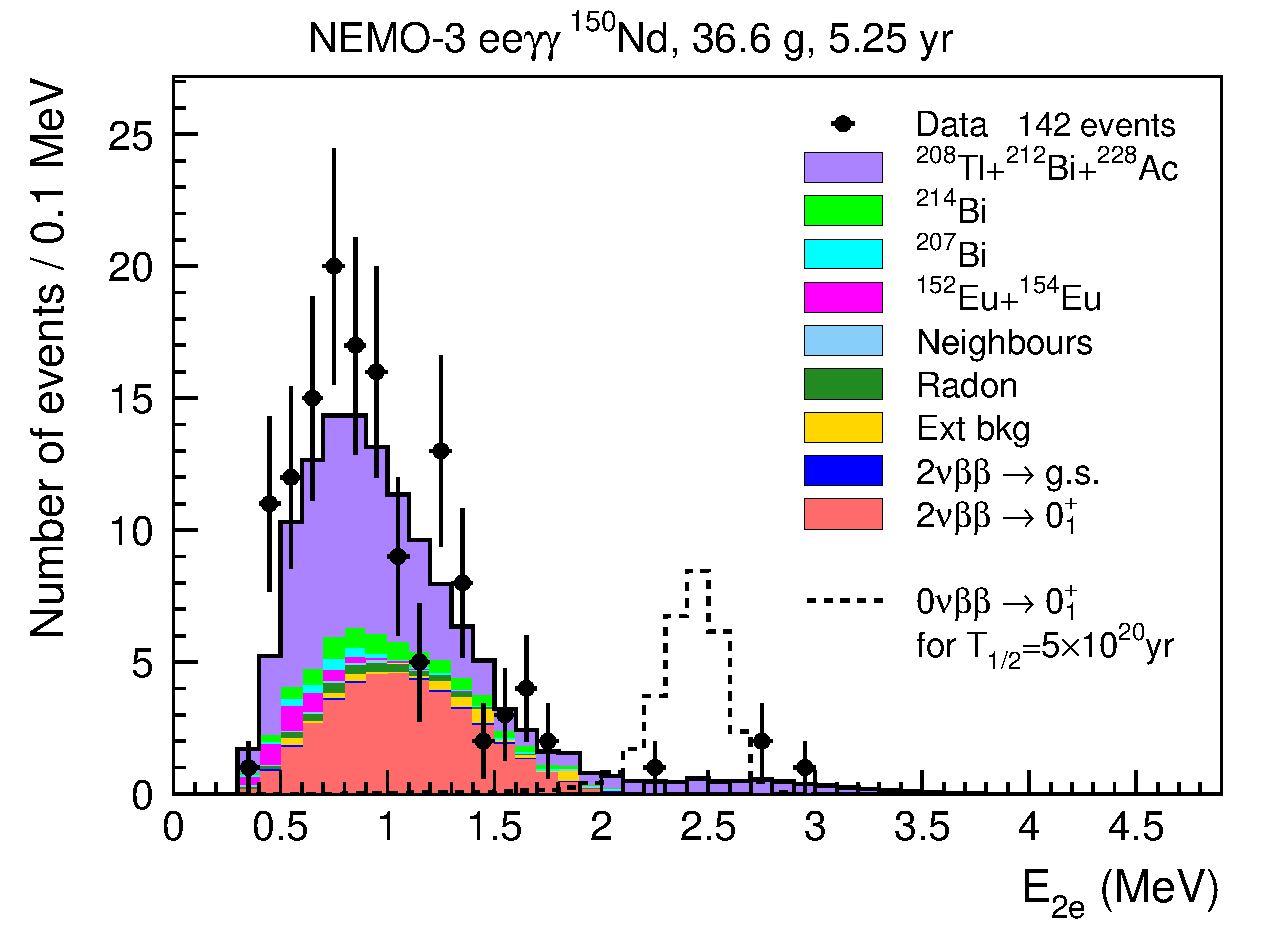}
\includegraphics[width=0.48\textwidth]{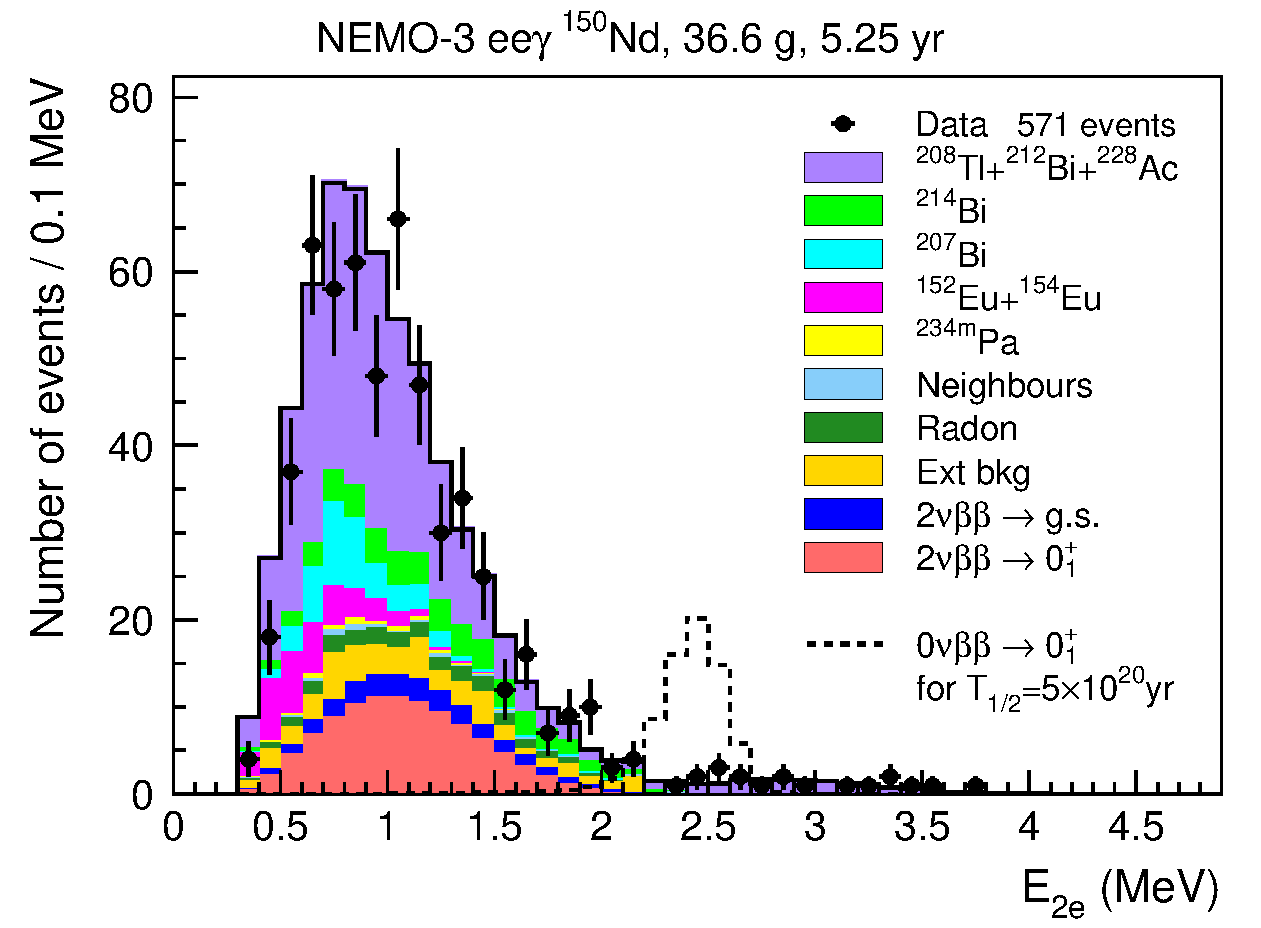}
\caption{Two-electron energy sum distributions for the $ee\gamma\gamma$ and $ee\gamma$ events  with the  superimposed 
signal expected for the 0$\nu\beta\beta$ decay to the 0$^+_1$ excited state}
\label{fig:e2e_0nu_m1}
\end{figure}
\begin{figure*}[h]
\includegraphics[width=0.48\textwidth]{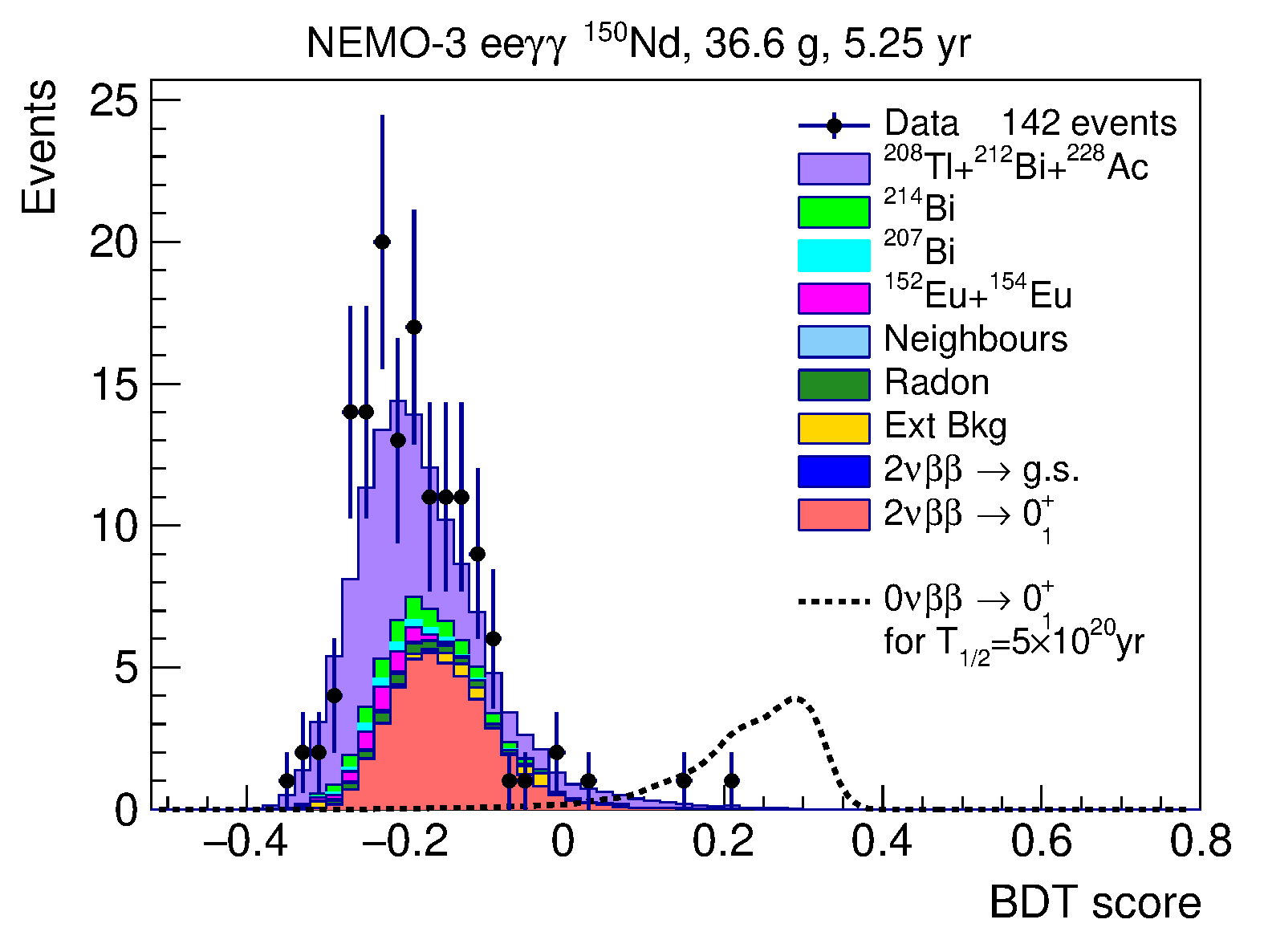}
\includegraphics[width=0.48\textwidth]{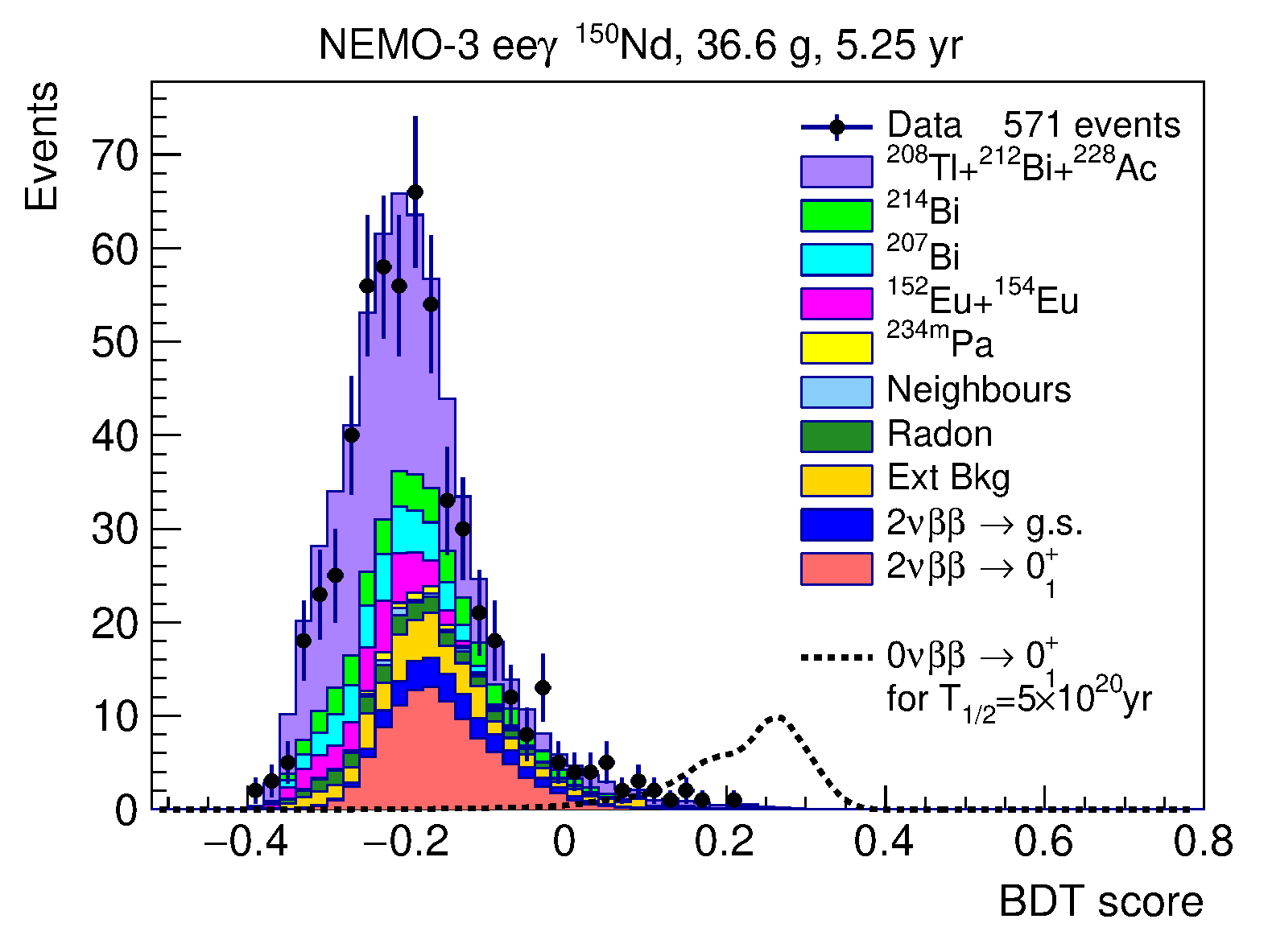}
\caption{BDT score distribution  for the $0\nu\beta\beta$ decay 
to the 0$^+_1$ excited state
in the $ee\gamma\gamma$ and $ee\gamma$ channels}
\label{fig:bdt_0nu_m1}
\end{figure*}
\subsection{Search for $\boldsymbol{0\nu\beta\beta}$ decay to $\boldsymbol{0^+_1}$ excited state}
The $ee\gamma\gamma$ and $ee\gamma$ events were used to search for the signal of the 0$\nu\beta\beta$ decay to the 0$^+_1$ excited state. 
No significant data excess is observed for the $0\nu\beta\beta$ signal in the distribution of the summed kinetic energy of two electrons (see Fig.~\ref{fig:e2e_0nu_m1}).
There are 4.2 (16.1) events expected in the region from 2~MeV to 2.8~MeV
under the 0$\nu\beta\beta$ peak, and 3 (16) data events are observed in the $ee\gamma\gamma$ ($ee\gamma$) channel.
The  BDT score distribution  is shown in Fig.~\ref{fig:bdt_0nu_m1}.

In the $ee\gamma\gamma$ channel, the observed half-life limit is
\begin{equation}
 T_{1/2}^{0\nu\beta\beta}(0^+_1) > 5.20 \times 10^{21}~\mbox{yr at 90\% C.L.} ~ ~ ,
\end{equation}
 and in the $ee\gamma$ channel it is
\begin{equation}
 T_{1/2}^{0\nu\beta\beta}(0^+_1) > 9.97 \times 10^{21}~\mbox{yr at 90\% C.L.}
\end{equation}
 When the combination of the two channels is used for calculating the limit, the result is 
\begin{equation} 
 T_{1/2}^{0\nu\beta\beta}(0^+_1) > 13.6 \times 10^{21}~\mbox{yr at 90\% C.L.}
\end{equation}
This limit is much stronger than that from \cite{Nasim}.
\begin{figure}[h]
\includegraphics[width=0.48\textwidth]{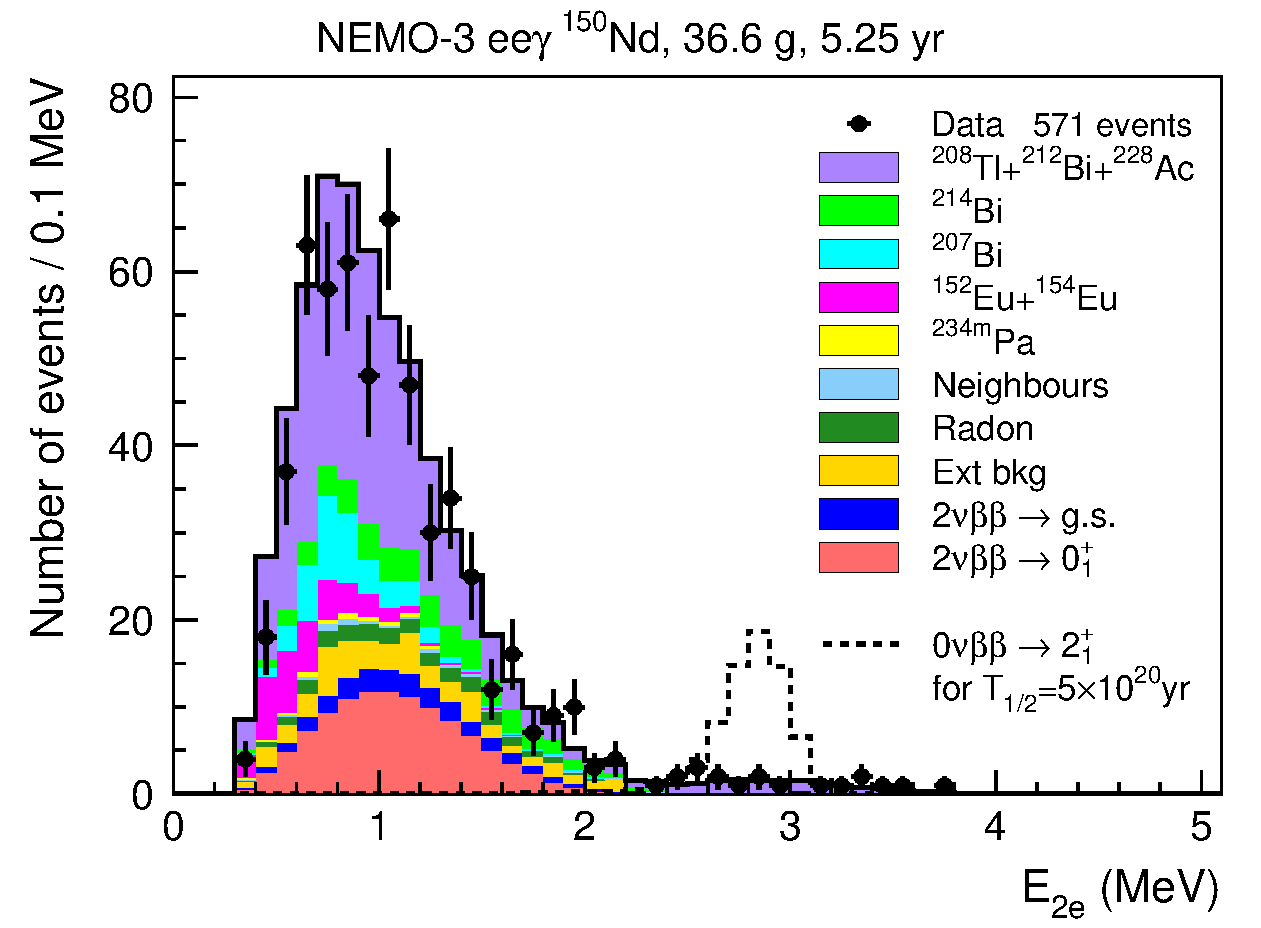}
\caption{Two-electron energy sum distribution for the $ee\gamma$ events with 
the signal expected for the 0$\nu\beta\beta$ decay to the 2$^+_1$ excited state superimposed}
\label{fig:e2e_eeg_0nu}
\end{figure}
\begin{figure}[h]
\includegraphics[width=0.48\textwidth]{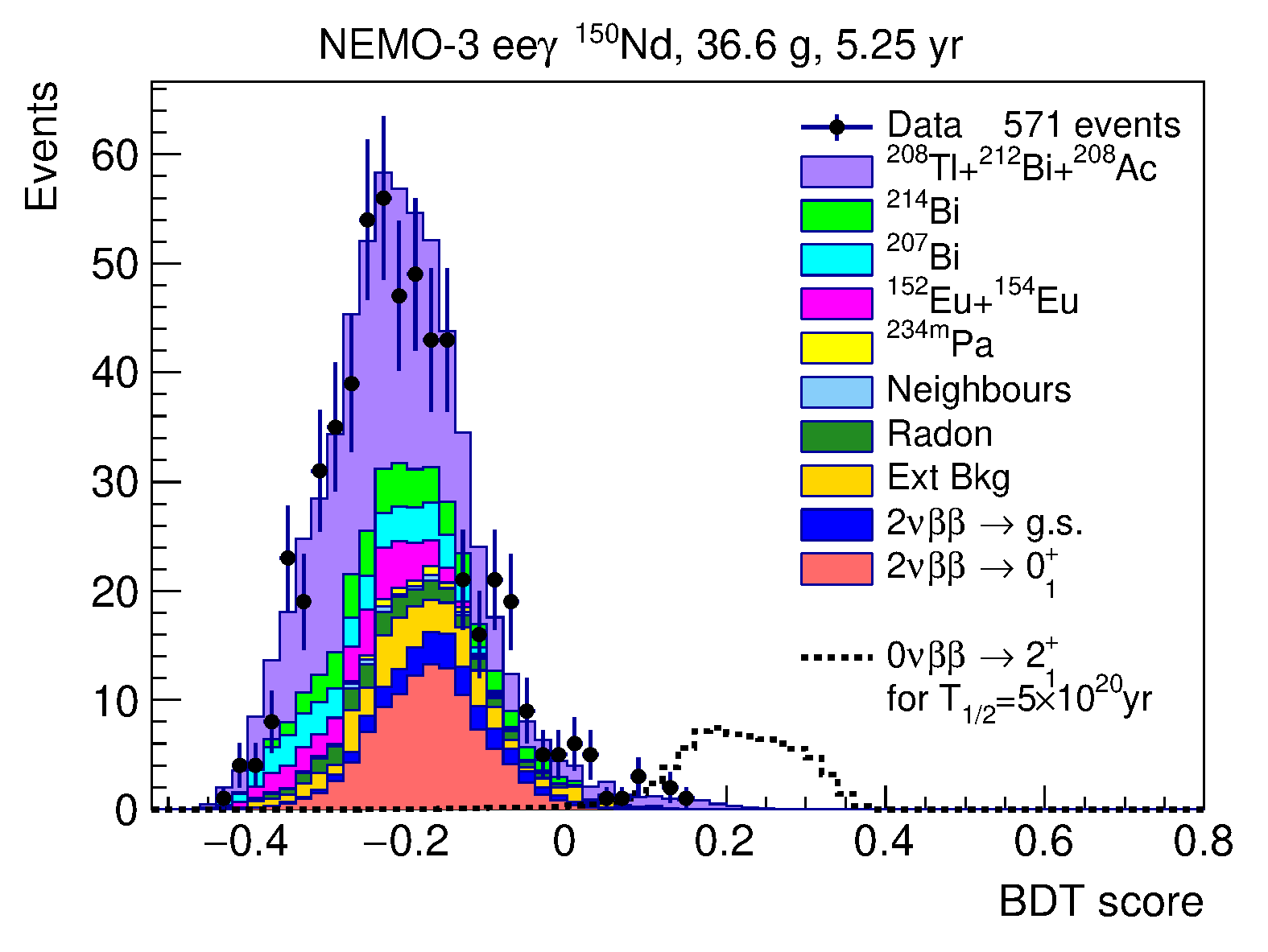}
\caption{BDT distribution for the
$0\nu\beta\beta$ decay to the 2$^+_1$ excited state}
\label{fig:bdt_eeg_0nu_2n}
\end{figure}
\subsection{Search for $\boldsymbol{0\nu\beta\beta}$ decay to $\boldsymbol{2^+_1}$ excited state}
The search for the $0\nu\beta\beta$ decay to the 2$^+_1$ excited state was performed using the
$ee\gamma$ events. The distribution of the summed kinetic energy of two electrons for these events is shown in Fig.~\ref{fig:e2e_eeg_0nu}. 
The 0$\nu\beta\beta$ signal in this distribution would be manifested in a peak located in the energy interval from 2.4~MeV to 3.2~MeV.
There are 12 data events in this interval in good agreement with the MC expectation of 11.6 events, meaning that no evidence of 
the $0\nu\beta\beta$ decay is observed.
The BDT score distribution for this decay mode is presented in Fig.~\ref{fig:bdt_eeg_0nu_2n}.
With this distribution, we obtain the following half-life limit:
\begin{equation}
T_{1/2}^{0\nu\beta\beta}(2^+_1) > 1.26 \times 10^{22}~\mbox{yr at 90\% C.L.}
\end{equation}
that is much more stringent than the result of  \cite{Nasim}, see Table~\ref{table:compare_t12}.
\section{Summary}
The two-neutrino double-$\beta$ decay of $^{150}$Nd to the 0$^+_1$ excited 
state of $^{150}$Sm has been measured with a statistical signal significance of 6.8 standard deviations and the signal-to-background ratio $S/B = 1.3$. 
The most precise measurement of the half-life for this process to date has been performed:
\begin{equation}
T_{1/2}^{2\nu\beta\beta}(0^+_1) = \left[ 1.11 ^{+0.19}_{-0.14} \,\left(\mbox{stat}\right) ^{+0.17}_{-0.15}\,\left(\mbox{syst}\right) \right] \times10^{20}~\mbox{yr}.
\end{equation}
The $\gamma\gamma$ angular correlation for this decay, measured for the first time, 
is found to be consistent with the expected behaviour characterising the
$0^{+} \to 2^{+} \to 0^{+}$  cascade.

No signal evidence was found for the $2\nu\beta\beta$ decay to the 2$^+_1$ excited state nor for the $0\nu\beta\beta$ decay to the 0$^+_1$ or 2$^+_1$ excited states. 
The corresponding 90\% confidence level limits have been established. The obtained half-life limit on the $2\nu\beta\beta$ decay to the 2$^+_1$ excited state 
\begin{equation}
T_{1/2}^{2\nu\beta\beta}(2^+_1) > 2.4 \times10^{20}~\mbox{yr at 90\% C.L.} 
\end{equation}
is slightly better than the best previous result
 of~\cite{PhysRevC.79.045501},
but still far short of the theoretical predictions
$T_{1/2} = 2.6\times10^{23}~\mbox{yr}$~\cite{Doi}, 
$T_{1/2} = 7.2\times10^{24}~\mbox{yr}$~\cite{Hirsch}.

The limits set on the neutrinoless decay half-life
have been significantly improved.
\section*{Acknowledgements}
We thank the staff of the Modane Underground Laboratory for their technical assistance
in running the experiment. We acknowledge support by the MEYS of the Czech Republic 
(Contract Number LM2023063), CNRS/IN2P3 in France, 
APVV in Slovakia (Projects No. 15-0576 and 21-0377),
NRFU in Ukraine (Grant No. 2020.02/0011),
STFC in the UK, and NSF in the USA.
\section*{Data Availability Statement}
The data that support the findings of this study are available upon reasonable request from the authors.


\begin{thebibliography}{10}
\providecommand{\url}[1]{{#1}}
\providecommand{\urlprefix}{URL }
\expandafter\ifx\csname urlstyle\endcsname\relax
  \providecommand{\doi}[1]{DOI \discretionary{}{}{}#1}\else
  \providecommand{\doi}{DOI \discretionary{}{}{}\begingroup
  \urlstyle{rm}\Url}\fi


\bibitem{Zyla:2020zbs}
R.L.~Workman, et~al. (Particle Data Group), Prog. Theor. Exp. Phys. \textbf{2022}(8), 083C01 (2022)

\bibitem{universe6100159}
A.S.~Barabash, Universe \textbf{6}(10), 159 (2020)

\bibitem{0nu-review}
M.~Agostini, et al., Rev. Mod. Phys. \textbf{95}, 025002 (2023)

\bibitem{Majorana}
E.~Majorana, Nuovo Cim \textbf{14}, 171 (1937)

\bibitem{Simkovic:2001ft}
F.~Simkovic, A.~Faessler, Prog. Part. Nucl. Phys. \textbf{48}, 201 (2002)

\bibitem{universe6120239}
P.~Belli, et~al., Universe \textbf{6}(12), 239 (2020)

\bibitem{10.3389/fphy.2019.00012}
S.~Stoica, M.~Mirea, Frontiers in Physics \textbf{7}, 12 (2019)

\bibitem{BarHubHub04}
A.S. Barabash, et~al., JETP Lett. \textbf{79}, 10 (2004)

\bibitem{PhysRevC.79.045501}
A.S. Barabash, et~al., Phys. Rev. C \textbf{79}, 045501 (2009)

\bibitem{PhysRevC.90.055501}
M.F. Kidd, et~al., Phys. Rev. C \textbf{90}, 055501 (2014)

\bibitem{Polischuk}
O.G. Polischuk, et~al., Phys. Scr. \textbf{96}, 085302 (2021)

\bibitem{Arnold:2016nd150}
R.~Arnold, et~al., Phys. Rev. D \textbf{94}, 072003 (2016)

\bibitem{Arnold:2004TDR}
R.~Arnold, et~al., Nucl. Instrum. Meth. A \textbf{536}, 79 (2005)

\bibitem{Arnold:2015wpy}
R.~Arnold, et~al., Phys. Rev. D \textbf{92}, 072011 (2015)

\bibitem{ROOT}
R.~Brun, F.~Rademakers, Nucl. Instrum. Meth. A \textbf{389}, 81 (1997)

\bibitem{TMVA}
A.~Hocker, et~al., CERN-OPEN-2007-007  (2007)

\bibitem{Junk:1999kv}
T.~Junk, Nucl. Instrum. Meth. A \textbf{434}, 435 (1999)

\bibitem{Read}
A.L. Read, J. Phys. G \textbf{28}, 2693 (2002)

\bibitem{Fisher:2006zz}
W.~Fisher, FERMILAB-TM-2386-E  (2006)

\bibitem{Ponkratenko:2000um}
O.A. Ponkratenko, V.I. Tretyak, Y.G. Zdesenko, Phys. Atom. Nucl. \textbf{63},
  1282 (2000)

\bibitem{Brun:1987GEANT3}
R.~Brun, et~al., CERN-DD-EE-84-1  (1987)

\bibitem{Evans}
R.D. Evans, \emph{The Atomic Nucleus} (McGrau-Hill inc., 1955)

\bibitem{Nasim}
J.~Argyriades, et~al., Phys. Rev. C \textbf{80}, 032501(R) (2009)

\bibitem{Stoica}
M.~Mirea, T.~Pahomi, S.~Stoica, Rom. Rep. Phys. \textbf{67}, 872 (2015)

\bibitem{nemo3-mo100-excited}
R.~Arnold, et~al., Nuclear Physics A \textbf{925}, 25 (2014)

\bibitem{Doi}
M.~Doi, T.~Kotani, E.~Takasugi, Progr. Theor. Phys. Suppl. \textbf{83}, 1
  (1985)

\bibitem{Hirsch}
J.~Hirsch, et~al., Nuclear Physics A \textbf{589}, 445 (1995)

\end{thebibliography}
\end{document}